%% file: eurosys25_main.tex
\documentclass[sigplan,twocolumn]{acmart}

\usepackage{amsmath,amssymb,amsfonts}
\usepackage{algorithm}
\usepackage[noend]{algpseudocode}
\usepackage{graphicx}
\usepackage{textcomp}
\usepackage{xcolor}
\usepackage{multirow}
\usepackage{lipsum}
\usepackage{subfigure}
\usepackage{url}
\usepackage{tabularray}
\usepackage{enumitem}
\usepackage{pifont}
\usepackage{caption}
\usepackage{boldline}
\usepackage{makecell}
\usepackage{titlesec}
\usepackage{balance}

\usepackage{listings} 

\lstdefinelanguage{json}{
    basicstyle=\normalfont\ttfamily,
    numberstyle=\scriptsize,
    stepnumber=1,
    numbersep=8pt,
    showstringspaces=false,
    breaklines=true,
    frame=lines,
    backgroundcolor=\color{gray!10},
    stringstyle=\color{blue},
    literate=
     *{:}{{{\color{purple}:}}}1
     {,}{{{\color{purple},}}}1
     {\{}{{{\color{brown}\{}}}1
     {\}}{{{\color{brown}\}}}}1
     {[}{{{\color{brown}[}}}1
     {]}{{{\color{brown}]}}}1,
}

\newcommand{\co}[1]{}  

\AtBeginDocument{%
  }

\copyrightyear{2025}
\acmYear{2025}
\setcopyright{acmlicensed}\acmConference[EuroSys '25]{Twentieth European Conference on Computer Systems}{March 30-April 3, 2025}{Rotterdam, Netherlands}
\acmBooktitle{Twentieth European Conference on Computer Systems (EuroSys '25), March 30-April 3, 2025, Rotterdam, Netherlands}
\acmDOI{10.1145/3689031.3717466}
\acmISBN{979-8-4007-1196-1/2025/03}

\acmSubmissionID{\#217}

\begin{document}


\newcommand{\occl}{\emph{DFCCL}}

\newcommand{\of}{OneFlow}

\newcommand{\Coll}{Collective}
\newcommand{\coll}{collective}

\newcommand{\Stick}{ Stickiness adjustment scheme}
\newcommand{\stick}{stickiness adjustment scheme}
\newcommand{\STICK}{\emph{Stickiness Adjustment Scheme}}

\newcommand{\Df}{Deadlock-free {\coll} execution framework}
\newcommand{\df}{deadlock-free {\coll} execution framework}
\newcommand{\DF}{\emph{Deadlock-free {\Coll} Execution Framework}}

\newcommand{\Dk}{Daemon kernel}
\newcommand{\dk}{daemon kernel}
\newcommand{\DK}{Daemon Kernel}

\newcommand\TODO[1]{\textcolor{blue}{\{#1\}}}

\title{Comprehensive Deadlock Prevention for GPU Collective Communication}

\author{Lichen Pan}
\orcid{0000-0001-7451-0140}
\affiliation{%
  \country{School of Computer Science, Peking University}
}
\email{plch368@pku.edu.cn}

\author{Juncheng Liu}
\affiliation{%
  \country{OneFlow Research}
}
\email{liujuncheng@oneflow.org}

\author{Yongquan Fu}
\affiliation{%
  \country{National Key Laboratory of  Parallel and Distributed Computing, College of Computer Science and Technology,
National University of Defense Technology}
}
\email{yongquanf@nudt.edu.cn}

\author{Jinhui Yuan}
\affiliation{%
  \country{OneFlow Research}
}
\email{yuanjinhui@oneflow.org}

\author{Rongkai Zhang}
\affiliation{%
  \country{School of Computer Science, Peking University}
}
\email{rkzhang@stu.pku.edu.cn}

\author{Pengze Li}
\affiliation{%
  \country{School of Computer Science, Peking University}
}
\email{lipengze@pku.edu.cn}

\author{Zhen Xiao}
\affiliation{%
  \country{School of Computer Science, Peking University}
}
\email{xiaozhen@pku.edu.cn}

\renewcommand{\shortauthors}{Lichen Pan et al.}

\begin{abstract}

Distributed deep neural network training necessitates efficient GPU collective communications, which are inherently susceptible to deadlocks.
GPU collective deadlocks arise easily in distributed deep learning applications when multiple collectives circularly wait for each other.
GPU collective deadlocks pose a significant challenge to the correct functioning and efficiency of distributed deep learning, and no general effective solutions are currently available.
Only in specific scenarios, ad-hoc methods, making an application invoke collectives in a consistent order across GPUs, can be used to prevent circular collective dependency and deadlocks.

This paper presents {\occl}, a novel GPU collective communication library that provides a comprehensive approach for GPU {\coll} deadlock prevention while maintaining high performance.
{\occl} achieves preemption for GPU {\coll}s at the bottom library level, effectively preventing deadlocks even if applications cause circular collective dependency.
{\occl} ensures high performance with its execution and scheduling methods for collectives.
Experiments show that {\occl} effectively prevents GPU {\coll} deadlocks in various situations.
Moreover, extensive evaluations demonstrate that {\occl} delivers performance comparable to or superior to NCCL, the state-of-the-art collective communication library highly optimized for NVIDIA GPUs.

\end{abstract}



\maketitle

\section{Introduction}
\label{sec:introduction}
\input{sections/introduction}

\section{Background and Motivations}
\label{sec:background}
\input{sections/background}

\section{Overview}
\label{sec:overview}
\input{sections/overview}

\section{Daemon Kernel}
\label{sec:dk}
\input{sections/daemonKernel}

\section{Implementation and Optimizations}
\label{sec:implement}
\input{sections/implement}

\section{Evaluation}
\label{sec:evaluation}
\input{sections/evaluation}

\section{Related Work}
\label{sec:related}
\input{sections/related}

\section{Conclusion}
\label{sec:conclusion}
\input{sections/conclusion}
\begin{acks}
The authors would like to thank the anonymous reviewers for their comments. This work was supported by the National Key R\&D Program of China under Grant 2023YFB2703800. The contact author is Zhen Xiao.
\end{acks}


\bibliographystyle{ACM-Reference-Format}
\bibliography{ref}

\clearpage
\appendix
\section{Artifact Appendix}

The artifact for the EuroSys 2025 paper, “Comprehensive Deadlock Prevention for GPU Collective Communication”, provides the code of {\occl} and the simulator mentioned in Sec. 2.4. The DOI of the artifact is at \url{https://doi.org/10.5281/zenodo.14871978}.

The code of {\occl} and the simulator is also publicly available at \url{https://github.com/Oneflow-Inc/dfccl}.

\end{document}

%% file: sections/introduction.tex
Recent years have witnessed the parameter count of deep neural network (DNN) models grow faster than the memory capacity and computational power of a single GPU ~\cite{gholami2020ai_and_memory_wall, para_count, ai_and_compute}.
%
This entails distributed DNN training, which includes various techniques such as data parallelism (DP)~\cite{2020vldb_pyrotch_ddp, sergeev2018horovod}, tensor parallelism (TP)~\cite{optimus2021, shoeybi2019megatron, bian2021_3d}, pipeline parallelism (PP)~\cite{huang2019gpipe_nips, narayanan2019pipedream_sosp, narayanan2021efficient_megatron}, and hybrid parallelism~\cite{narayanan2021efficient_megatron, bian2021_3d, rasley2020deepspeed}.
GPU {\coll} communications are used to synchronize DNN status and play a critical role in distributed DNN training.

\begin{figure*}[!t]
    \centering

    \subfigure[\textbf{Legal}: consistent order.]{
        \begin{minipage}[t]{0.235\textwidth}
            \centering
            \includegraphics[width=\textwidth]{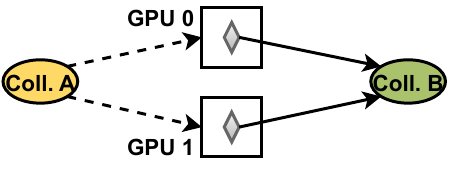}
            \label{fig:normal_1}
        \end{minipage}    
    }
    \subfigure[\textbf{Legal}: disorder with sufficient resources.]{
        \begin{minipage}[t]{0.235\textwidth}
            \centering
            \includegraphics[width=\textwidth]{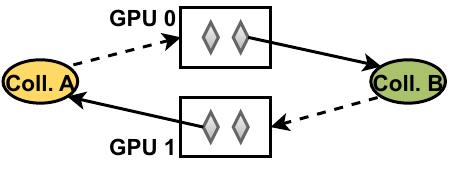}
            \label{fig:normal_2}
        \end{minipage}    
    }
    \subfigure[\textbf{Deadlock}: disorder with single queue or resource-depletion.]{
        \begin{minipage}[t]{0.235\textwidth}
            \centering
            \includegraphics[width=\textwidth]{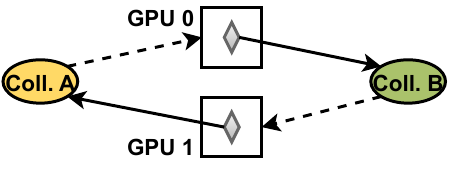}
            \label{fig:nccl_deadlock_1}
        \end{minipage}
    }
    \subfigure[\textbf{Deadlock}: disorder with GPU synchronization, despite sufficient resources.]{
        \begin{minipage}[t]{0.235\textwidth}
            \centering
            \includegraphics[width=\textwidth]{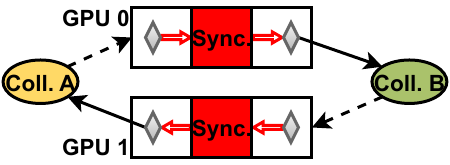}
            \label{fig:nccl_deadlock_2}
        \end{minipage}
    }
    \caption{\textbf{Legal Situations and Basic Deadlock Situations.} \textnormal{In the figures, both collective A and collective B execute on GPU 0 and GPU 1. The diamonds within each GPU represent the resource "units" required to execute a collective on that GPU, assuming collective A and B require the same amount of resources. Solid arrows indicate that a resource unit has been allocated to a collective, which means the collective is \textit{executing} on that GPU. Dashed arrows represent a collective applying for a resource unit from a GPU, signifying that the collective is \textit{invoked} on that GPU but is not executing. Outlined red arrows depict the dependencies of idle resources on allocated resources introduced by GPU synchronization.}}
    \label{fig:nccl_deadlock}
\end{figure*}

Widely used GPU {\coll}s are vulnerable to deadlocks, because they work in a resource-holding, busy-waiting way, and preemption is ill-supported on GPUs.
%
GPU collective deadlocks occur when an application causes circular collective dependency.
We summarize the \textit{basic deadlock situations} of GPU collectives (Fig.~\ref{fig:nccl_deadlock}) at the bottom library level of the distributed deep learning stack.
The \textit{disordered invocation of {\coll}s} across different GPUs emerges as a necessary condition for circular collective dependency and deadlocks, while \textit{GPU synchronization} (see Sec.~\ref{sec:back_basic}) exacerbates the circular collective dependency.

We conduct simulation experiments to quantitatively analyze the susceptibility of GPU collectives to deadlocks.
%
%
The simulation experiments (Table~\ref{table:sim}) show that low probabilities of {\coll} disorder and GPU synchronization (both 0.004\%) can cause high deadlock risks (6.94\%), and the deadlock ratio is more sensitive to GPU synchronization than to disordered collective invocation.

GPU collective deadlocks pose significant threats to the correct functioning, training efficiency, and hardware utilization of distributed deep learning.
%
GPU {\coll} deadlocks can cause GPUs to show 100\% utilization with no progress being made, leading to wasted resources~\cite{torch_issue_31095, torch_issue_52142, torch_issue_24081, tf_issue_44976}.
%
The deadlocks are challenging to identify and resolve with little error messages or logs clearly indicating the problem~\cite{checkpoint-2024jiangNSDI2024megascale, checkFreq}.
%

Existing methods fail to deal comprehensively and effectively with GPU collective deadlocks.
%
%
Current methods to prevent circular {\coll} dependency and deadlocks are stopgap case-by-case solutions that make an application invoke collectives in a consistent order across GPUs, with GPU synchronization unmanaged.
At present, manual hardcoding, which is expensive to develop and verify, and tightly coupled with applications, remains the sole practical approach when integrating PP with other parallel techniques.
In more complex and irregular distributed DNN training scenarios~\cite{barham2022pathways, chowdhery2023palm}, applying the labor-intensive yet ad hoc manual collective orchestration method is quite challenging, and uncontrolled GPU synchronization further diminishes its effectiveness.
%

This paper presents {\occl} (Deadlock Free Collective Communication Library), which is, to the best of our knowledge, the first GPU {\coll} communication library that provides a comprehensive approach for GPU {\coll} deadlock prevention and maintains high performance.

{\occl} offers a general and effective method to prevent deadlocks in various scenarios via \textit{preemption}, fundamentally breaking the inherent susceptibility of GPU {\coll}s to deadlocks.
{\occl} executes {\coll}s in a \textit{two-phase blocking} manner~\cite{feitelson1992gang, ousterhout1982scheduling}  in the \textit{{\dk}}, preempting {\coll}s deemed to be \textit{stuck} via context switch.
%
With preemptive support at the bottom library level, circular collective dependency from applications no longer causes deadlocks.

{\occl} maintains high performance with three key designs:
%
\ding{182} {\occl} employs the on-GPU control logic and the busy-waiting~\cite{nccl} execution mode before preempting a {\coll} to ensure high throughput.
\ding{183} {\occl} makes each GPU independently perform {\coll} preemption in a \textit{decentralized dynamic} manner to avoid coordination overheads among GPUs.
%
\ding{184} {\occl}'s adaptive scheduling scheme, which supports user-specified priority and achieves decentralized dynamic gang-scheduling for {\coll}s, ensures efficient scheduling and execution of collectives.

We implement {\occl} based on NVIDIA GPUs.
%
%
{\occl} can be seamlessly integrated with existing distributed deep learning frameworks~\cite{yuan2021oneflow, paszke2019pytorch, abadi2016tensorflow} by appropriately substituting their NCCL API calls with {\occl} API calls.

Experimental results show that {\occl} effectively prevents GPU {\coll} deadlocks.
We compare {\coll} bandwidth and latency between {\occl} and NCCL~\cite{nccl}, the state-of-the-art (SOTA) collective communication library highly optimized for NVIDIA GPUs.
%
We conduct experiments to compare the DNN training performance using {\coll}s from {\occl} versus NCCL across various distributed training techniques and frameworks.
%
Evaluation results show that {\occl} delivers performance comparable to NCCL, which requires various scenario-dedicated CPU orchestration, and achieves performance gains under certain circumstances.

This paper makes the following contributions:

\noindent $\bullet$ We quantify the analysis of deadlocks and the influencing factors through simulation experiments.

\noindent $\bullet$ We identify the preemption chances for GPU {\coll}s and introduce {\occl}, a novel GPU {\coll} communication library that offers a comprehensive approach to preventing GPU {\coll} deadlocks with performance guarantees.

\noindent $\bullet$ Experimental results show that {\occl} effectively prevents GPU {\coll} deadlocks and achieves performance comparable to or superior to NCCL.


%% file: sections/background.tex
\begin{table*}[h]
\centering
\caption{Configurations and Deadlock Ratios in Simulation Based Analysis.}

\label{table:sim}
\resizebox{1\textwidth}{!}{
\begin{tabular}{cV{2.5}c|c|c|c|c|c|cV{2.5}c|c|c|c|c|c}
\hlineB{2.5}

\multirow{2}{*}{} & \multicolumn{7}{cV{2.5}}{\textbf{3D Grouping Policy}} & \multicolumn{6}{c}{\textbf{Free Grouping Policy}} \\ \Xcline{2-14}{0.8pt}

 &
{\shortstack{TP, DP, PP\\Group Size}} &
{\shortstack{\#Group}} &
{\#GPU} & {\shortstack{\#{\Coll} /\\TP, DP Group}} &
{\shortstack{Disorder\\Prob.}} & {\shortstack{Sync.\\Prob.}} &
{\shortstack{\textbf{Deadlock}\\\textbf{Ratio}}} &
{\shortstack{\#Group}} &
{\#GPU} & {\shortstack{\#{\Coll} /\\Group}} &
{\shortstack{Disorder\\Prob.}} &
{\shortstack{Sync.\\Prob.}} &
{\shortstack{\textbf{Deadlock}\\\textbf{Ratio}}} \\ \hlineB{2.5}
 
\multirow{4}{*}{\shortstack{\textbf{Single-}\\\textbf{Queue}\\\textbf{Model}}} & 
\multirow{2}{*}{4, 4, 4} &
\multirow{2}{*}{32} &
\multirow{2}{*}{64} & 
\multirow{2}{*}{400, 1200} & 1e-7 & - & \textbf{1.10\%} & 1 & 8 & 161 & 1e-5 & - & \textbf{1.21\%} \\ \cline{6-14}

 &  &  &  &  & 1e-6 & - & \textbf{9.97\%} &
\multirow{2}{*}{32} &
\multirow{2}{*}{64} &
\multirow{2}{*}{400 or 1200} & 1e-6 & - & \textbf{0.98\%} \\ \cline{2-8} \cline{12-14}
 
 & 
\multirow{2}{*}{8, 6, 64} &
\multirow{2}{*}{896} &
\multirow{2}{*}{3072} &
\multirow{2}{*}{400, 1200} & 1e-9 & - & \textbf{0.47\%} &  &  &  & 1e-5 & - & \textbf{9.45\%} \\ \cline{6-14}

 &  &  &  &  & 1e-8 & - & \textbf{3.59\%} & 32 & 128 & 400 or 1200 & 1e-6 & - & \textbf{1.72\%} \\ \hlineB{2.5}

\multirow{5}{*}{\shortstack{\textbf{Sync.}\\\textbf{Model}}} &
\multirow{4}{*}{4, 4, 4} &
\multirow{4}{*}{32} &
\multirow{4}{*}{64} &
\multirow{3}{*}{400, 1200} & 2e-3 & 4e-3 & \textbf{0.68\%} &
\multirow{4}{*}{32} &
\multirow{4}{*}{64} &
\multirow{3}{*}{400 or 1200} & 4e-6 & 4e-5 & \textbf{0.81\%} \\ \cline{6-8} \cline{12-14}

 &  &  &  &  & 4e-3 & 4e-3 & \textbf{1.38\%} &  &  &  & 4e-5 & 4e-5 & \textbf{1.16\%} \\ \cline{6-8}  \cline{12-14}

 &  &  &  &  & 4e-3 & 2e-3 & \textbf{0.32\%} &  &  &  & 4e-5 & 8e-5 & \textbf{6.56\%} \\ \cline{5-8} \cline{11-14}
 
 &  &  &  & 800, 2400 & 4e-3 & 4e-3 & \textbf{2.56\%} &  &  & 800 or 2400 & 4e-5 & 4e-5 & \textbf{6.94\%} \\ \cline{2-14}
 
 & 8, 6, 64 & 896 & 3072 & 400, 1200 & 8e-4 & 8e-4 & \textbf{1.56\%} &
32 & 128 & 400 or 1200 & 4e-5 & 4e-5 & \textbf{2.34\%} \\ \hlineB{2.5}
\end{tabular}
}
\end{table*}

\subsection{NCCL}
\label{sec:back_nccl}

NCCL provides GPU {\coll}s efficiently utilizing inter-GPU bandwidth, and is commonly used in distributed DNN training.
%
%
Experiments show that the throughput of NCCL all-reduce surpasses that of CUDA-aware MPI~\cite{nv_cuda_mpi, mvapich} when the buffer size exceeds 32 KB.
The increase in throughput reaches over 6.7$\times$ at most.

NCCL boosts throughput by shifting from the CPU-based control plane to the on-GPU, busy-waiting control logic, which, however, makes NCCL collectives deadlock-prone.

\subsection{Community Discussions}
\label{sec:back_community}

The documentation of fundamental distributed DNN training tools merely alerts developers to the risk of NCCL deadlocks without providing practical solutions~\cite{nccl_deadlock, nccl_mpi_deadlock, nccl_iommu, pytorch_warning, amd_iommu}.

We investigate issues reporting NCCL-related deadlocks in Pytorch, DeepSpeed, and TensorFlow repositories.
When a deadlock occurs, typical symptoms include the program hanging, GPU utilization stuck at 100\%, NCCL logs failing to provide valuable information.
%
Despite extensive discussions, no final solutions are proposed in many issues, while in others, only ad-hoc workarounds are offered, such as introducing fences or turning off certain hardware features.

The deadlocks reported in some issues conform to the GPU synchronization-related deadlock case (Sec.~\ref{sec:back_basic}),
and the improvised methods avoid the mutual interference between GPU synchronization and {\coll}s.
Issue \#31095 reported in PyTorch~\cite{torch_issue_31095} reports a scenario where after the invocation of GPU {\coll}s, a process gets stuck at the code line responsible for page-locked host memory allocation.
When a  sleep interval is added between the GPU {\coll}s and the memory allocation, this problem disappears.
%
Issues reporting NCCL deadlocks from various repositories~\cite{torch_issue_52142, torch_issue_24081, tf_issue_44976} suggest disabling IOMMU~(Input-Output Memory Management Unit), 
which triggers CPU-initiated GPU memory operations, leading to implicit GPU synchronization.

\subsection{Analysis of GPU Collective Deadlocks}
\label{sec:back_basic}

NCCL {\coll}s are vulnerable to deadlocks since they inherently satisfy three out of the four individually necessary and jointly sufficient conditions for a deadlock (except \textit{circular waiting}):
%
\ding{182} \textbf{\textit{Mutual exclusion}}: GPU resources, e.g., streaming multiprocessors and shared memory~\cite{CUDA_guide}, occupied by one collective cannot be simultaneously used by others.
\ding{183} \textbf{\textit{Hold and wait}}:
In NCCL, the parts of a {\coll} on different GPUs busy-wait until all peers are ready while holding allocated resources.
%
\ding{184} \textbf{\textit{No preemption}}: There is no practical official preemption support for GPUs, and the GPU-preemption techniques in literature
are not suitable for {\coll}s (see Sec.~\ref{sec:related}).
Therefore, once an application causes circular collective dependency, a deadlock occurs.

At the bottom library level, Fig.~\ref{fig:nccl_deadlock} shows the legal situations and the \textbf{basic deadlock situations} of GPU collectives.

When an application invokes collectives in a consistent order on each GPU, e.g., invoking collective B before A on both GPU 0 and 1 (Fig.~\ref{fig:normal_1}), these collectives can execute normally.
CUDA stream~\cite{cuda_stream} enables the parallel execution of multiple collective kernels when resources are sufficient.
%
When an application invokes collectives in different orders on each participating GPU, if these collectives are issued to different streams, and the kernels in all these streams can be scheduled for execution due to sufficient resources (Fig.~\ref{fig:normal_2}), these collectives can execute normally.

Fig.~\ref{fig:nccl_deadlock_1} and \ref{fig:nccl_deadlock_2} summarize three basic GPU collective deadlock situations with circular collective dependency at bottom library level.
\ding{182} \textbf{Single Queue}: For the single queue programming model where collectives are issued in a single stream on each GPU, the disordered invocation of {\coll}s on different GPUs leads to circular collective dependency and a deadlock (Fig.~\ref{fig:nccl_deadlock_1}).
%
%
\ding{183} \textbf{Resource Depletion}: When collectives are issued to different streams, but resources are insufficient, the disordered invocation of {\coll}s on different GPUs leads to the deadlock situation similar to Single Queue (Fig.~\ref{fig:nccl_deadlock_1}).
%
\ding{184} \textbf{GPU Synchronization Related}: 
GPU synchronization is categorized into explicit and implicit types~\cite{CUDA_guide}.
%
Explicit synchronization involves \textsl{\textsf{cudaDeviceSynchronize()}} calls, while implicit synchronization includes GPU default stream commands, page-locked host memory allocation, and CPU-initiated GPU memory operations.
%
GPU synchronization operations block and suspend a GPU until all kernels in all streams of the GPU complete.
%
Therefore, GPU synchronization introduces the dependency where idle resources cannot be allocated to collectives invoked after the GPU synchronization, until all allocated resources are released by prior collectives after their completion.
%
As shown in Fig.~\ref{fig:nccl_deadlock_2}, when two GPUs invoke two collectives in opposite orders, and both issue GPU synchronization after invoking one, the disordered collective invocation and the resource dependency introduced by GPU synchronization together lead to circular collective dependency and a deadlock.
%

Disordered {\coll} invocation across GPUs is a necessary condition for circular collective dependency and deadlocks, with GPU synchronization exacerbating the risk of circular waiting among collectives.
The root cause of the disordered {\coll} invocation and the issuance of GPU synchronization originates with the application.
%
In applications where collectives lack data dependency, these collectives can be invoked in varying orders across GPUs.
GPU synchronization, independent of these collectives, can also be issued whenever needed on corresponding GPUs.
%


\subsection{Simulation Based Analysis}
\label{sec:back_sim}

\subsubsection{Deadlock Simulator}
We develop a simulator to quantitatively analyze how disordered collective invocation and GPU synchronization affect GPU {\coll} deadlocks.

Our simulation is driven by real-world distributed DNN training practices and profiling~\cite{shoeybi2019megatron, narayanan2021efficient_megatron, yuan2021oneflow}.
GPUs are organized into different \textit{groups}.
%
Each group has a separate {\coll} list for its GPUs to \textit{invoke} and \textit{execute}.
A single GPU can belong to multiple groups, which means that the {\coll}s a GPU will invoke and execute are the union of the {\coll}s from all the groups the GPU belongs to.
%
A collective has three states on each GPU: \textit{invoked}, \textit{executing}, and \textit{successful}. 
%
A collective becomes successful when it reaches the executing state on all GPUs in its corresponding group.
%
The condition for a collective's transition from invoked to executing state is related to the \textbf{deadlock decision model}.

We study two deadlock decision models in the simulator based on the basic deadlock situations discussed in Sec.~\ref{sec:back_basic}:

\noindent $\bullet$ \textbf{Single-queue model.}
A collective on a GPU transitions to the executing state if there is no executing or invoked collectives before it.
Each GPU is restricted to having only one executing {\coll} at a time.
%
%

\noindent $\bullet$ \textbf{Synchronization model.}
A GPU may randomly initiates synchronization operations that suspends it.
A collective on a GPU transitions to the executing state if it is invoked before the GPU is suspended.
A GPU ends suspension when all executing collectives before the synchronization transition to successful, or if there are no executing collectives before the synchronization.
Each GPU can maintain an unlimited number of executing {\coll}s.
%
Given the varying total resources of different GPU models and the diverse resource needs of {\coll}s, the simulator employs an idealized infinite resource assumption to simplify these complexities.

\begin{figure}[tb]
  \centering
  \includegraphics[width=0.35\textwidth]{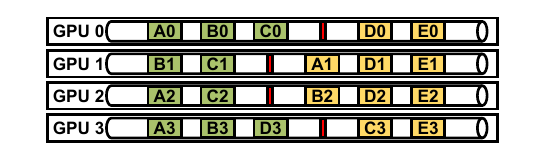}
  \caption{\textbf{Example Collective States in the Synchronization Model.} \textnormal{A, B, C, D, and E are five collectives. A\textit{i} represents the part of collective A on GPU \textit{i}. The left green box indicates a collective is \textit{executing} on a GPU, and each right yellow box indicates an \textit{invoked} collective.
  The middle red bar represents the synchronization that suspends a GPU.}}
  \label{fig:sim}
\end{figure}

In both deadlock decision models, the simulator checks for cycles in the \textit{dependency graph} to determine the presence of a deadlock, after each collective invocation and synchronization issuance.
The \textit{nodes} of the dependency graph are collective parts on GPUs. 
The graph includes two types of \textit{directed dependency edges}: \ding{182} an executing collective on one GPU points to all its invoked counterparts on other GPUs, e.g., D3->D0, D3->D1, and D3->D2 in Fig.~\ref{fig:sim}; \ding{183} an invoked collective on a GPU points to all executing collectives on the same GPU, e.g., D0->A0, D0->B0, and D0->C0 in Fig.~\ref{fig:sim}.
One of the cycles in the dependency graph corresponding to Fig.~\ref{fig:sim} is A0->A1->B1->B2->C2->C3->D3->D0->A0.

The simulation models key behaviors related to GPU {\coll} deadlocks:
(1) the \textit{disorder probability} specifies the probability of disordered {\coll} invocations by GPUs; (2) the  \textit{synchronization probability}
determines the probability of issuing synchronization operations.

\begin{figure}[tb]
  \centering
  \includegraphics[width=0.38\textwidth]{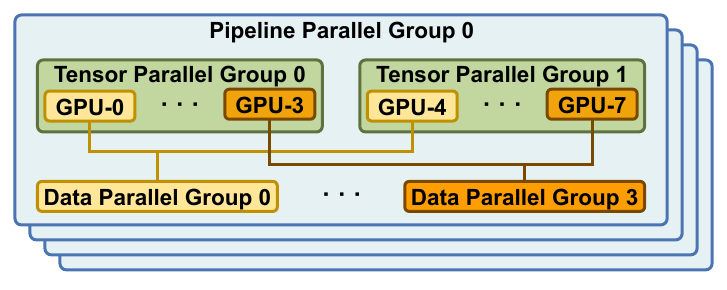}
  \caption{Example Grouping of GPUs in 4-way Tensor, 2-way Data, and 4-way Pipeline Hybrid Parallelism~\cite{shoeybi2019megatron, narayanan2021efficient_megatron}.}
  \label{fig:pp}
\end{figure}

We study two typical \textbf{GPU grouping policies} for distributed DNN training:

\noindent $\bullet$ \textbf{3D grouping policy}.
As shown in Fig.~\ref{fig:pp}, GPUs are organized according to the 3D grouping scheme in 3D-hybrid parallel distributed DNN training~\cite{shoeybi2019megatron, narayanan2021efficient_megatron}, where GPUs holding the same DNN model part in different TP groups form a DP group within each PP group.
The configuration file specifies the sizes of the TP, DP, and PP groups, along with the number of {\coll}s in the TP and DP groups.

\noindent $\bullet$ \textbf{Free grouping policy.}
The configuration file directly specifies the total number of groups, as well as the GPU lists and the number of {\coll}s of each group.

The simulator's input is synthesized event sequences for each GPU. These sequences, consisting of {\coll} invocation events and GPU synchronization events, are generated based on disorder probability, synchronization probability, and GPU grouping.
The simulator transitions the {\coll} state according to currently submitted events and determines if deadlocks occur using the deadlock decision model.

\subsubsection{Simulation Setup}
\label{sec:back_sim_setup}

Table~\ref{table:sim} summarizes the configurations and deadlock ratios of the simulation experiments.

\noindent $\bullet$ The (8, 6, 64)-3D grouping case is inspired by the training configuration of GPT-3~\cite{narayanan2021efficient_megatron}.
The (1, 8)-free grouping case simulates a data parallel scenario.

\noindent $\bullet$ In 3D grouping, each GPU invokes {\coll}s from two groups.
In contrast, in the (32, 64)-free grouping case, which mirrors the (4, 4, 4)-3D grouping in total groups and GPUs, GPUs variably receive {\coll}s from one to five groups.

\noindent $\bullet$ 
In the (32, 64)-free grouping case, 28 groups have three GPUs each, and four groups have eight GPUs each.
The (32, 128)-free grouping case increases each group by two GPUs.

\noindent $\bullet$ 
In the free grouping policy, "400 or 1200" means 50\% of groups have 400 collectives and the other 50\% have 1200 collectives.
"800 or 2400" preserves this distribution.

\noindent $\bullet$ A \textit{round} is defined as all {\coll}s are successful or the simulation runs until a deadlock arises.
For all the configurations, deadlock ratios are calculated from 32,000 rounds.

\subsubsection{Result Analysis}
\label{sec:back_sim_result}

From Table~\ref{table:sim}, we can conclude:

\noindent \ding{182} Extremely low disorder and synchronization probabilities can lead to prohibitively high deadlock risks.
In the single-queue model, the deadlock ratio is up to six orders of magnitude higher than the disorder probability (0.49\% vs. 1e-9). In the synchronization model, the deadlock ratio is up to three orders of magnitude higher than the disorder and synchronization probability (6.94\% vs. 4e-5).

\noindent \ding{183} The deadlock ratio is positively correlated with the disorder probability and the synchronization probability.

\noindent \ding{184} 
The synchronization model is more sensitive to the synchronization probability than to the disorder probability.
%
For example, in the (32, 64)-free grouping case, while increasing the disorder probability tenfold (from 4e-6 to 4e-5) raises the deadlock ratio by 42\%, doubling the synchronization probability (from 4e-5 to 8e-5) dramatically increases the deadlock ratio by 468\%.
%
This is because GPU synchronization introduces the dependency of idle resources on allocated resources, which facilitates the emergence of circular waiting among disordered collectives,
and results in deadlocks
even with ample resources (Fig.\ref{fig:nccl_deadlock_2} vs. Fig.\ref{fig:normal_2}).

\noindent \ding{185} The deadlock ratio correlates positively with the total number of GPUs and the total number of planned {\coll}s.

\noindent \ding{186} 
In the synchronization model, the deadlock ratio is positively correlated with the group overlapping degree, i.e., the number of groups a GPU belongs to, which indicates the complexity of distributed DNN training scenarios.
%
For example, the (32, 64)-free case employs disorder and synchronization probabilities two orders of magnitude lower than the (4, 4, 4)-3D case (4e-5 vs. 4e-3), yet their reported deadlock ratios are similar (1.16\% vs. 1.38\%).

Sec.~\ref{sec:eval_deadlock} demonstrates that in a real-world environment, when both disorder and synchronization probabilities are engineered to reach 100\%, NCCL exhibits a 100\% deadlock ratio, whereas {\occl} does not encounter any deadlocks.

\subsection{Existing Methods to Deal With Deadlocks}
\label{howExistworks}

Existing methods prevent circular {\coll} dependency by ensuring that collectives are invoked in a consistent order across GPUs, without managing GPU synchronization.
%
Most approaches achieve this through additional CPU orchestration tightly coupled with specific parallel training techniques.
Manual hardcoding is currently the only viable solution when combining PP with other parallel techniques.

Different CPU coordination strategies are employed for data parallelism.
\ding{182} Horovod~\cite{sergeev2018horovod} presents the dynamic centralized coordinating approach.
%
The Horovod central coordinator gathers {\coll}s' readiness from each GPU during runtime and broadcasts a list of {\coll}s ready on all GPUs, allowing GPUs to start the all-reduces in the list order. 
\ding{183}
BytePS~\cite{jiang2020unified_byteps} requires centralized coordination prior to invoking {\coll}s among intra-node GPUs.
\ding{184}
KungFu~\cite{2020MaiOSDI20KungFu} determines the predominant GPU collective calling order in the initial training step via gather and broadcast operations. Subsequently, decentralized schedulers enforce this order across all GPUs.
\ding{185}
%
{\of}~\cite{yuan2021oneflow} introduces a static-sorting based scheduling approach. Its compiler automatically constructs task graphs for all GPUs, sorting {\coll}s based on each graph's topological order. During runtime, GPUs initiate {\coll}s following these pre-sorted sequences.

Horovod, BytePS, and KungFu are incapable of orchestrating all collectives in 3D-hybrid parallelism.
\ding{182}
Megatron-LM~\cite{shoeybi2019megatron, narayanan2021efficient_megatron} introduces manual hardcoding for hybrid parallelism by manually and meticulously arranging each GPU's {\coll}s related to different groups (Fig.~\ref{fig:pp}).
\ding{183} OneFlow follows the manual hardcoding scheme when combining PP with other parallel techniques.

\subsubsection{Limitations of Existing Methods}
Case-by-case collective orchestration at the application level is neither effective nor general in preventing GPU collective deadlocks.

Manual hardcoding is indeed an expensive ad hoc method.
%
The existing manual orchestration implementation demands extensive development and verification by experienced engineers, and is closely tied to hybrid parallelism characteristics.

%
As distributed training patterns grow more complex and dynamic, it is becoming increasingly challenging for engineers to manually orchestrate collectives at the application level, ensuring that GPUs invoke them in the consistent order under all runtime circumstances.
%
Pathways~\cite{barham2022pathways, chowdhery2023palm} presents a heterogeneous and dynamic distributed training paradigm less symmetrical and more irregular than 3D-hybrid parallelism\footnote{Pathways relies on the centralized scheduler and a closed-source dataflow system called PLAQUE to ensure that all participating devices invoke collectives in a consistent order. The specific implementation details are not publicly available. However, as described in its paper, Pathways also relies on additional CPU orchestration to prevent deadlocks in device collectives, making it fundamentally indistinguishable from existing approaches.}.
This paradigm is conceptually similar to the (32, 64)-free grouping case in Sec.~\ref{sec:back_sim_setup}.
Table~\ref{table:sim} shows, in the (32, 64)-free grouping case, disorder and synchronization probabilities of no more than 0.004\% yield a deadlock risk near 7\%.
%
Besides, Sec.~\ref{sec:back_sim_result} shows deadlocks are more sensitive to GPU synchronization than to disordered collective invocation.
%
This implies that in dynamic, complex scenarios, as long as the disorder probability is not zero, laborious {\coll} orchestration can be futile in preventing deadlocks due to uncontrolled synchronization fluctuations, e.g., more frequent GPU memory allocations.




%% file: sections/overview.tex
\begin{figure}[tb]
  \centering
\includegraphics[width=0.43\textwidth]{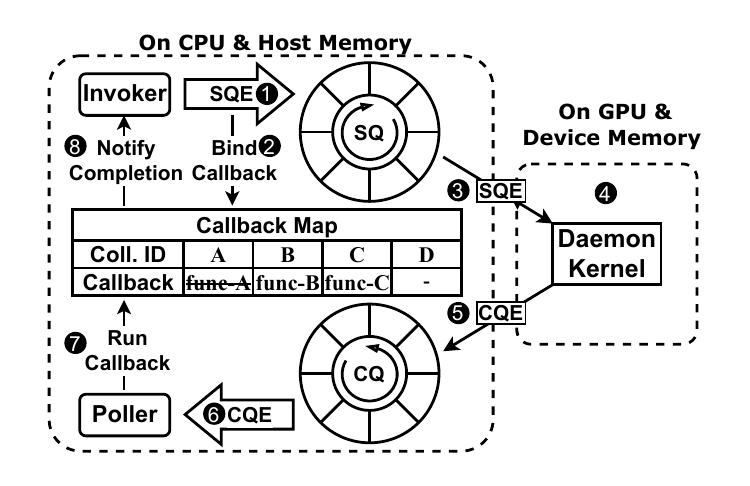}
  \caption{\textbf{Overview of {\occl}.} \textnormal{A, B, C, and D denote four registered {\coll}s. A is complete; B and C are invoked; and D has not been invoked yet.}}
  \label{fig:overview}
\end{figure}

{\occl} is a GPU collective communication library that prevents deadlock  through preemption, and maintains high performance with efficient execution and scheduling.

\subsection{{\occl} Components and Collective Life Cycle}

{\occl} consists of GPU and CPU parts, as shown in Fig.~\ref{fig:overview}.
On each GPU, the \textit{{\dk}} (Sec.~\ref{sec:dk}), {\occl}'s core component, handles execution, preemption, and scheduling for multiple {\coll}s.
{\occl}'s CPU component provides user-friendly APIs and manages asynchronous, non-blocking request submitting and completion notifying based on the \textit{submission queue (SQ)} and \textit{completion queue (CQ)}.
Each GPU's {\dk} corresponds to a separate SQ \& CQ.

Fig.~\ref{fig:overview} identifies a {\coll}'s life cycle in {\occl} through numerical labels.
\noindent \ding{182}, \ding{183}: The {\coll} invoker inserts a \textit{submission queue element (SQE)} into the SQ, and records a ({\coll} ID, callback) pair in the callback map.
%
\noindent \ding{184}, \ding{185}: The {\dk} periodically checks the SQ, fetches and parses SQEs, and executes the requested {\coll}s.
\noindent \ding{186}: The daemon kernel inserts a \textit{completion queue entry (CQE)} for each completed {\coll} into the CQ.
\noindent \ding{187}, \ding{188}: The \textit{poller} thread monitors the CQ.
Once the poller finds a CQE, it executes the callback tied to the {\coll}, which notifies the invoker of the {\coll}’s completion in the user-defined way.

Using callbacks for {\coll}s' asynchronous invocation is a common programming practice~\cite{paszke2019pytorch, yuan2021oneflow}.
{\occl}'s design offers a flat learning curve.

\subsection{APIs of {\occl}}
\label{sec:overview_api}

Listing~\ref{lst:interface} provides a list of {\occl}'s APIs.
\textsl{\textsf{\textbf{dfcclInit}}} initializes the rank context of a GPU.
\textsl{\textsf{\textbf{dfcclDestroy}}} destroys the rank context and releases resources.
\textsl{\textsf{\textbf{dfcclRegister*}}} (``*'' represents a specific {\coll}) registers a {\coll} and prepares corresponding data structures for a {\coll} on the specified GPU.
Every registered {\coll} has a unique ID.
\textsl{\textsf{\textbf{dfcclRun*}}} invokes a registered {\coll} based on the ID and records a user-defined callback corresponded to the {\coll}.


In {\occl}, a {\coll} is registered once with \textsl{\textsf{dfcclRegister*}} and can then be invoked repeatedly as needed using \textsl{\textsf{dfcclRun*}}. {\occl} also allows dynamic registration of new {\coll}s during runtime.

The \textit{communicator} manages the resources for inter-GPU data transfer during {\coll} execution.
{\occl} maintains a communicator pool transparent to users, automatically creating and allocating communicators for {\coll}s.

{\occl} can be seamlessly integrated with existing distributed deep learning frameworks, e.g, PyTorch~\cite{paszke2019pytorch}, OneFlow~\cite{yuan2021oneflow}, TensorFlow~\cite{abadi2016tensorflow}, by appropriately substituting their calls to the NCCL APIs with calls to the {\occl} APIs.


\begin{figure}[tb] 
\begin{small}
\noindent\begin{minipage}{\linewidth}
\begin{lstlisting}[language=C++, caption={\textbf{APIs of {\occl}.} Take all-reduce as an example.}, label=lst:interface]
ret_t dfcclInit(rankCtx_t* rankCtx, int rank);
ret_t dfcclRegisterAllReduce(size_t count, type_t type, redOp_t op, int collId, devSet_t devSet, int priority, rankCtx_t rankCtx);
ret_t dfcclRunAllReduce(const void* sendbuff, void* recvbuff, int collId, func_t callback, void* callbackArgs, rankCtx_t rankCtx);
ret_t dfcclDestroy(rankCtx_t rankCtx);
\end{lstlisting}
\end{minipage}
\end{small}
\end{figure}

\subsection{Benefits of {\occl}}

%
{\occl}'s daemon kernel integrates preemptive scheduling on GPU, fundamentally breaking GPU collectives' inherent susceptibility to deadlocks.
%
This eliminates the need for CPU-based coordination between user-facing interfaces and DFCCL's library-level APIs (e.g., \textsl{\textsf{dfcclRunAllReduce}}).
%
%
Embedding scheduling logic in a daemon kernel rather than relying on CPU orchestration is novel and has the potential to be extrapolated to other CPU-GPU paradigms.

%
{\occl} achieves three key objectives simultaneously: 1) the APIs of the underlying collective communication library can be directly called concurrently and asynchronously; 2) guaranteeing high performance; and 3) maintaining independence from specific parallel training techniques, thus ensuring wide applicability.
The combined realization of these three objectives represents an open research challenge that existing work has not addressed.

\textbf{{\occl} vs. NCCL}
The methodology difference between {\occl} and NCCL includes two main aspects.

\noindent \ding{182} {\occl} manages the execution, preemption, and scheduling for an arbitrary number of collectives submitted dynamically in a single daemon kernel.
In contrast, each NCCL kernel is dedicated to one or a few predetermined collectives, relying entirely on CUDA's underlying scheduling.

\noindent \ding{183}
{\occl} offers SQ, CQ, and callback management for asynchronous request submitting and completion notifying, unlike NCCL, which requires additional mechanisms, e.g., cudaEvent~\cite{CUDA_guide}, to verify {\coll} completion asynchronously.

%% file: sections/daemonKernel.tex
In this section, we first present the preemption chances for GPU  {\coll}s.
Next, we present {\coll} preemption and scheduling in the {\dk}.
Then, we present the {\dk}'s voluntary quitting and event-driven starting.
Finally, we analyze its correctness and performance.



\subsection{Preemption Chance of GPU {\Coll}s}
\label{sec:dk_chance}

The preemption opportunity for common GPU collectives (all-reduce, all-gather, reduce-scatter, reduce, and broadcast) arises because they are all composed of a subset of the same group of \textit{primitives}~\cite{cowan2023msccl,nccl}.
%
In a {\coll}, GPUs are organized into a specific logical topology, with each GPU assigned a primitive sequence based on its position within this topology.
To facilitate processing, input data for a {\coll} are divided into regular \textit{chunks}.
GPUs execute a collective by performing its primitive sequence a certain number of times to process all the data chunks.

\begin{figure}[tb]
  \centering
  \includegraphics[width=0.32\textwidth]{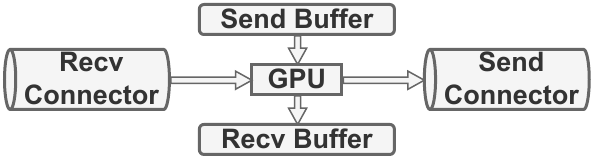}
  \caption{Buffers Used in GPU Collectives.}
  \label{fig:buffer}
\end{figure}

Every primitive is a fusion of basic actions, i.e., \textsl{\textsf{send}}, \textsl{\textsf{recv}}, \textsl{\textsf{reduce}}, and \textsl{\textsf{copy}}, which describe the basic operations on the four buffers used in {\coll}s as shown in Fig.~\ref{fig:buffer}.
\textsl{\textsf{send/recv buffers}} are local buffers for input and output.
\textsl{\textsf{send/recv connectors}} contain lock-free ring buffers used for inter-GPU data transfer, managed by the \textit{communicator}.
%
The \textsl{\textsf{send}} action writes data to the \textsl{\textsf{send connector}}, while \textsl{\textsf{recv}} reads data from the \textsl{\textsf{recv connector}}.
%
The \textsl{\textsf{reduce}} action reduces data from the \textsl{\textsf{send buffer}} and the \textsl{\textsf{recv connector}} with a specified \textit{reducing function}.
The \textsl{\textsf{copy}} action puts data into the \textsl{\textsf{recv buffer}}.
%
Each primitive includes one or both of the \textsl{\textsf{send}} or \textsf{\textsl{recv}} actions.
%
Based on the presence of \textsl{\textsf{send}} and \textsl{\textsf{recv}} actions, a primitive busy-waits until the \textsl{\textsf{send connector}} is writable and/or the \textsl{\textsf{recv connector}} is readable before progressing.
%

%

By limiting a primitive's wait time, we can abort its execution, thereby \textit{preempting} the associated {\coll}.
Moreover, the \textit{persistent visibility} of written data enables individual GPUs to independently preempt {\coll}s in a \textit{decentralized dynamic} manner, without explicit coordination among GPUs.
%
%
Once a GPU writes data to the \textsf{\textsl{send connector}} for a collective's primitive, the data remain visible to the peer GPU. 
The visibility persists even if the collective part on this GPU is preempted after writing, or the corresponding collective part on the peer GPU is preempted before writing.
%

\subsection{{\Coll} Execution and Preemption}
\label{sec:dk_preemption}

\begin{figure}[tb]
  \centering
  \includegraphics[width=0.45\textwidth]{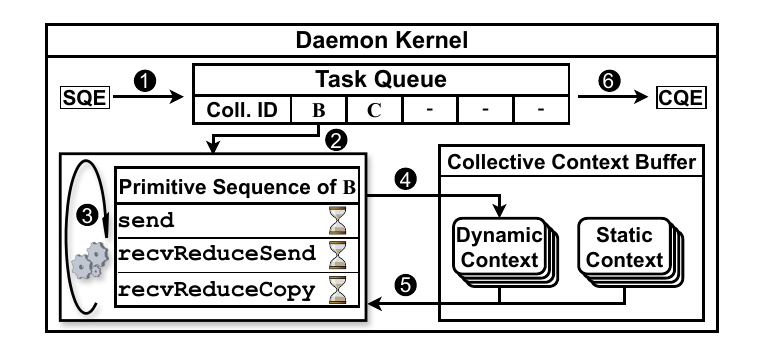}
  \caption{\textbf{{\DK} of {\occl}}. \textnormal{{\Coll} B and C are invoked, and B is scheduled currently.}}
  \label{fig:dk}
\end{figure}

%
Fig.~\ref{fig:dk} demonstrates key components of the {\dk} and identifies operations related to {\coll} execution and preemption through numerical labels. 

\noindent \ding{182}: The {\dk} fetches and parses SQEs, and maintains {\coll}s in the \textit{task queue} in the shared memory~\cite{CUDA_guide}.

\noindent \ding{183}: The {\dk} traverses the task queue and schedules a {\coll}, details of scheduling are discussed in Sec.~\ref{sec:dk_gang}.

\noindent \ding{184}: The {\dk} executes the primitive sequence of the scheduled {\coll} in a \textit{two-phase blocking} manner~\cite{feitelson1992gang, ousterhout1982scheduling}.
%
The {\dk} assigns \textit{spin thresholds} to the {\coll}'s primitives to limit the busy-waiting time.
During execution, a primitive first polls up to spin-threshold times to check if the condition required by the \textsl{\textsf{send}} and/or \textsl{\textsf{recv}} action is met.
If the primitive cannot execute after polling spin-threshold times, it is aborted, and the associated {\coll} is deemed to be \textit{stuck} and then preempted on this GPU.
%

\noindent \ding{185}, \ding{186}:
The \textit{context} of the preempted {\coll} is saved in the \emph{{\coll} context buffer} in the global memory~\cite{CUDA_guide}, and the {\dk} loads the next scheduled {\coll}'s context into the \textit{active context slot} in the shared memory.
The context of {\coll}s consists of \emph{dynamic context} and \emph{static context}.
%
The dynamic context includes changing states during {\coll} execution, e.g., the current data chunk ID and the aborted primitive's ID.
%
The static context of a {\coll} contains its constant configuration, such as local and \textsl{\textsf{connector}} buffers' addresses shown in Fig.\ref{fig:buffer}, and its meta information including the number of GPUs executing the {\coll}, the rank of the current GPU among participants, and the composition of the {\coll}'s primitive sequence, etc.
%
%
Before executing a {\coll}, the {\dk} loads both its dynamic and static context, yet only saving its dynamic context after preemption, as the static context remains unchanged during {\coll} execution.
However, the {\coll}'s static context can change across multiple calls, e.g., the addresses of \textsl{\textsf{send buffer}} and \textsl{\textsf{recv buffer}} may vary.

\noindent \ding{187}: The {\dk} inserts a CQE into the CQ for the completed {\coll}.

\subsection{Adaptive {\Coll} Scheduling}
\label{sec:dk_gang}

Algorithm~\ref{algo:scheduling} shows {\occl}'s scheduling process.
{\occl}'s {\dk} schedules {\coll}s via the \emph{adaptive {\stick}}, which enables priority assigning and decentralized dynamic gang-scheduling for {\coll}s.

The \emph{stickiness} of a {\coll} indicates the {\dk}'s willingness to wait for its progress.
%
A {\coll}'s stickiness is reflected in its position in the task queue and the spin thresholds assigned to its primitives.
The unified stickiness adjustment mechanism independently controls both aspects on each GPU, supporting various policies.
All the GPUs adopt the same stickiness adjustment policy, which is further divided into an \textit{order adjusting policy} and a \textit{spin threshold adjusting policy}.
The order adjusting policy controls the frequency of fetching SQEs from the SQ and the ordering of {\coll}s in the task queue, according to user-specified priorities (line~\ref{algo:line_ordering}).
%
The spin threshold adjusting policy enables each GPU to negotiate in a decentralized dynamic manner to execute the same {\coll} (line~\ref{algo:line_profile} and line~\ref{algo:line_threshold}).

\begin{algorithm}[tb]
\begin{small}
\caption{Scheduling Process of {\occl}.}
\label{algo:scheduling}
\textbf{with} \textsl{\textsf{Pointer2SQ}}, \textsl{\textsf{SpinThreshold}}, \textsl{\textsf{PrimitiveExecuteStatus}}, and \textsl{\textsf{Pointer2CQ}} \textbf{in} \textsl{shared memory}
\begin{algorithmic}[1]  
\While{not $\textsc{FinallyExit()}$}
    \State $\textsc{FetchSQE}\&\textsc{SortTaskQueueByPriority}()$\label{algo:line_ordering}
    \State $\textsc{SetInitialSpinThreshold()}$\label{algo:line_profile}
    \For{$\text{collective} \in \text{task queue}$}
        \State $\textsc{LoadCollectiveContext()}$
        \For{$\text{primitive} \in \text{collective}$}
            \State $\textsc{ExecutePrimitive}()$
            \If{$\textsc{PrimitiveSuccess}()$}
                \State $\textsc{AdaptivelyAdjustSpinThreshold}()$\label{algo:line_threshold}
            \Else
                \State $\textsc{PreemptCollective}()$
                \State $\textsc{SaveCollectiveContext()}$
                \State \textbf{break}
            \EndIf
        \EndFor
        \If{$\textsc{CollectiveSuccess()}$}
            \State $\textsc{SendCQE()}$
        \EndIf
    \EndFor
\EndWhile
\end{algorithmic}
\end{small}
\end{algorithm}

%
When no specific priority is assigned to {\coll}s, the {\dk} adopts a FIFO-ordering policy; otherwise, it employs a priority-based ordering policy.

\noindent $\bullet$ \textbf{FIFO Ordering.} This policy aims to empty the task queue quickly.
%
Under this policy, the {\dk} fetches an SQE from the SQ when the task queue is empty or all {\coll}s in the queue cannot progress for a while.
The newly fetched {\coll} is put at the end of the task queue. 

\noindent $\bullet$ \textbf{Priority-based Ordering.} This policy optimizes performance for applications' special scheduling needs.
Under this policy, the {\dk} checks the SQ more frequently and sorts the task queue by priority.
A practical priority scheme is assigning higher priority to {\coll}s arriving later to enable the overlap of communication and computation in data parallelism~\cite{peng2019generic_bytescheduler, bao2020preemptive_pace, 2020vldb_pyrotch_ddp}.

When a collective is preempted, it remains at its original task queue position, and the {\dk} schedules the next collective in the task queue for execution.

{\occl}'s automated, adaptive spin threshold adjusting policy, which enables GPUs to negotiate in a decentralized dynamic manner to achieve gang-scheduling, plays a pivotal role in forestalling inter-GPU conflicts.
This policy instructs the {\dk} to assign the largest initial spin threshold to the {\coll} at the front of the task queue.
%
Each subsequent {\coll} receives a progressively lower initial spin threshold (line~\ref{algo:line_profile}).
During execution, if the {\dk} detects that a primitive of a {\coll} executes successfully, it raises the spin threshold of the succeeding primitives of that {\coll} (line~\ref{algo:line_threshold}), which increases the probability that all GPUs simultaneously execute or wait for the same {\coll}, resulting in de facto gang-scheduling.

{\occl} uses automated profiling to set suitable parameters, e.g., the initial spin threshold and voluntary quitting period (Sec.~\ref{sec:dk_voluntary}), to achieve the Pareto-optimal (Sec.~\ref{sec:dk_performance}).


\subsection{Voluntary Quitting and Event-driven Starting}
\label{sec:dk_voluntary}

The {\dk} voluntarily quits when it cannot fetch new SQEs for a certain period and the task queue is either empty or contains only {\coll}s that cannot progress.
The {\dk} voluntarily quits for two reasons: 

\noindent $\bullet$ To release GPU resources for other tasks when it's idle.

\noindent $\bullet$ To prevent deadlocks related to GPU synchronization.
Once the {\dk} quits, the blocking GPU synchronization can complete, thus allowing stuck {\coll}s to proceed.

{\occl} tries to start the {\dk} upon SQE insertion to the SQ or when the inserted CQEs are fewer than SQEs.
%
The {\dk} is initially launched upon the insertion of the first SQE.
%
\textsl{\textsf{dfcclDestroy}} inserts an \emph{exiting SQE} into the SQ, making the {\dk} \textit{finally exit} after reading it.

\subsection{Correctness and Performance Analysis}
\label{sec:dk_performance}

\indent \textbf{Correctness.}
The {\dk} ensures the correctness of dynamic, decentralized preempting and restoring of {\coll}s, by preserving the context integrity of preempted, uncompleted {\coll}s.
%
Saving and loading the preempted {\coll}'s dynamic context ensures it restarts from the previous stopping point, without under- or re-transmitting data.
%
The {\dk} prevents other {\coll}s from using preempted, uncompleted {\coll}'s \textsl{\textsf{connectors}}, ensuring the correct exploitation of the data visibility (Sec.~\ref{sec:dk_chance}).
Besides, the {\dk}'s voluntary quitting and restarting do not corrupt preempted {\coll}s' context in global memory.

\textbf{Performance Modeling.}
The overheads $T$ in {\coll} execution includes the busy-waiting time ($t[spin]$), the context switch time ($t[switch]$), and the waiting time for scheduling influenced by task queue length ($t[q\_len]$).
\begin{equation}
  T =   t[spin] + t[switch] + t[q\_len] \label{eq:total}
\end{equation}
%
$t[spin]$ correlates positively with the spin threshold ($N_{spin}$).
%
Both $t[switch]$ and $t[q\_len]$ are negatively correlated with $N_{spin}$: a larger $N_{spin}$ increases the probability that a {\coll} successfully waits for the same {\coll}'s scheduling on peer GPUs, so it experiences fewer preemptions and context switches, and completes faster, thereby reducing the task queue length.
%
Therefore, we can assess the correlation between the overheads $T$ and the spin threshold $N_{spin}$ via expression~\ref{eq:sim}.
In {\occl}, adaptively adjusting the spin threshold, as a uniform approach, is used to approximate a Pareto-optimal~\cite{deb2011pareto} for overheads in various scenarios.
\begin{align}
    T \sim N_{spin} + \frac{1}{N_{spin}} \label{eq:sim}
\end{align}

\subsection{Discussions}

{\occl} introduces at least two innovations for deadlock-free GPU {\coll}s:
%
\ding{182} A daemon kernel that achieves two-phase blocking execution of GPU {\coll}s
on the hardware platform without preemption support, 
without altering the underlying GPU task scheduling mechanism.
%
\ding{183} A {\coll} preemption and scheduling co-design that enables decentralized, dynamic collaboration of multiple GPUs without explicit coordination among them, enhancing traditional timer-based context switch in preemptive scheduling~\cite{andrew2015modern}.


During the two-phase blocking execution, {\occl} can preempt a collective at any time by interrupting any primitive.
%
The specific method of interrupting a primitive involves assigning an appropriate spin threshold before its execution, causing it to yield if no progress is made within the spin threshold.
In contrast, in NCCL, a primitive busy-waits indefinitely while holding resources.

%% file: sections/implement.tex
\textit{\textbf{Implementation Details of the {\DK}}}
We tailor the {\dk} to the block-thread programming model of CUDA~\cite{CUDA_guide}.
A CUDA kernel comprises multiple parallel threads executing the same code.
The threads are grouped into blocks.
All threads of a block reside in the same streaming multiprocessor (SM) and have access to the SM's limited shared memory.
Equally-shaped blocks further form a grid.
Each {\coll} is assigned a specific grid and block sizes.
The {\dk} is launched using the largest grid and block sizes among all registered {\coll}s.

Threads within the same block synchronize easily, while those in different blocks typically run asynchronously.
%
When the {\dk} executes {\coll}s requiring different numbers of blocks simultaneously, its higher-index blocks can execute a different {\coll} than lower-index blocks.
%
%
The number of active threads executing a {\coll}'s primitives inside a block depends on the block size assigned to the {\coll}. The {\dk} makes extra threads wait.
Each block independently decides when to quit voluntarily.
%
%

The {\dk}'s reads from the SQ and writes to the CQ are implemented to accommodate asynchronous block execution.
%
The SQ is a single-producer-multi-consumer (SPMC) ring buffer, allowing only one CPU thread to write an SQE at a time.
All blocks of the {\dk} read the SQE.
When a block reads a new SQE, it atomically increases a counter inside the SQE.
%
If a block finds the increased counter of an SQE equals the {\dk}'s grid size, it marks the corresponding SQ slot as writable.
%
A block executes a {\coll} only if its index is lower than the {\coll}'s assigned grid size.
%
The CQ is a multi-producer-single-consumer (MPSC) ring buffer.
Only one poller thread on the CPU reads CQEs.
The {\dk} maintains a completion counter for each {\coll} in global memory.
A block atomically increases the {\coll}'s completion counter when completing its part of a {\coll}.
If a block finds the increased completion counter of a {\coll} equals its assigned grid size, the block writes a CQE to the CQ and resets the {\coll}'s completion counter.
Multiple blocks can concurrently write CQEs into the CQ for different completed {\coll}s.

The CPU \& GPU cache-coherence mechanism is transparent to SQ/CQ management, and we use memory fences for memory consistency when necessary.
We use CUDA's atomic APIs that ensure transparent atomic read/write operations between GPUs and DRAM, as well as within a GPU.

\textit{\textbf{Optimizing CQ.}}
The SQ and CQ reside in page-locked host memory.
We reduce host-memory-related reads, writes and memory fences when writing CQEs to decrease CQE-writing latency by leveraging CUDA's 64-bit atomic operations.
The \textbf{vanilla ring-buffer-based CQ} requires at least five host-memory-related operations to prevent blocks from inserting CQEs for different {\coll}s into the same CQ slot.
%
The vanilla ring-buffer-based CQ also requires a memory fence between writing the CQE and updating the CQ's \textsf{\textsl{tail}} to ensure memory consistency.
%
The \textbf{optimized ring-buffer-based CQ} uses exactly four host-memory-related operations without fences by encapsulating the complete {\coll}'s ID and current \textsl{\textsf{tail}} in a single 64-bit atomic write.
The poller validates a CQE by comparing CQ's \textsl{\textsf{head}} and the \textsl{\textsf{tail}} from the 64-bit bitmap.
%
We further develop an \textbf{optimized CQ} that only requires at least a single CUDA \textsl{\textsf{atomicCAS\_system}} operation to write a CQE, abandoning the ring buffer semantics.
This optimized CQ is based on the observation that the CQE only carries the complete {\coll}'s ID.
A block atomically writes the ID into a writable slot in the CQ.
%
The poller scans the CQ, checking whether a slot contains a valid {\coll} ID and marking the slot writable after reading the {\coll} ID.
SQ is implemented as a vanilla ring buffer because an SQE contains more information than a 64-bit word can hold.

\textit{\textbf{Reducing the Overheads of Context Switching.}}
We employ three methods to minimize context-switching overheads.
\ding{182}
The {\dk} loads and saves the context in parallel with multiple threads.
We encapsulate the dynamic and static context into 16-byte aligned \textsf{\textsl{structs}} to utilize the 16-byte \textsf{\textsl{load/store}} instructions.
\ding{183}
The {\dk} uses multiple active context slots in shared memory, managed with a direct-mapped cache approach.
\ding{184}
The {\dk} employs a lazy-saving strategy, only saving the dynamic context of a {\coll} that has progressed before preemption.


\textit{\textbf{Integrating {\occl} with Frameworks.}}
%
We extend {\of}~\cite{yuan2021oneflow} as well as PyTorch~\cite{paszke2019pytorch} \& Megatron-LM~\cite{shoeybi2019megatron, checkpoint-2024jiangNSDI2024megascale} to use {\occl}-based {\coll}s. The integration with each framework requires approximately 1,000 lines of C++ code to invoke proper {\occl} APIs.

%% file: sections/evaluation.tex
\begin{table}[tb]
  \centering
  \caption{Specifications of the Experimental Platforms.}
  \label{table:specs}
  \resizebox{0.45\textwidth}{!}{
  \begin{tabular}{c | c }
    \hline
                      & Specification \\
    \hline
    Processor         & Intel Xeon Silver 4314 @ 2.40GHz (16 cores $\times$ 2 sockets) \\ 
    DRAM              & 512GB @ 2666 MT/s \\
    GPU               & NVIDIA GeForce RTX 3080 Ti 12GB $\times$ 8 \\     
    NIC                & Mellanox MT28908 @ 56Gb/s \\
    \hline
    Processor         & Intel Xeon Silver 4314 @ 2.40GHz (16 cores $\times$ 2 sockets) \\ 
    DRAM              & 512GB @ 2666 MT/s \\
    GPU               & NVIDIA GeForce RTX 3090 24GB $\times$ 8 \\     
    NIC                & Mellanox MT28908 @ 56Gb/s \\    
    \hline
    Switch              & Mellanox SX6036 (36 full-duplex 56Gb/s ports) \\
    \hline
  \end{tabular}
  }
\end{table}


In this section, we conduct experiments to verify {\occl}'s deadlock prevention capability and measure its performance.


\noindent {$\bullet$ \textbf{Testbed.}}
We conduct experiments on the platforms detailed in Table~\ref{table:specs}, using Ubuntu 20.04 and CUDA 11.7.
On the dual-socket servers, GPUs 0-3 and GPUs 4-7 belong to two separate \textsl{\textsf{PIX}} domains, and these two device groups reside within the \textsl{\textsf{SYS}} domain.
%
GPUs within each machine communicate via the Shared Memory (SHM) transports, while inter-machine communication utilizes RDMA networking.
These machines are hereafter referred to as the 3080ti-server and the 3090-server.
The primitive sequences for {\coll}s are generated with Simple protocol and Ring algorithm~\cite{nccl}.

\noindent {$\bullet$ \textbf{Benchmarks.}} 
%
\ding{182} Verifying {\occl}'s deadlock-prevention capability.
\ding{183} Measuring {\occl}'s workload-independent overheads.
\ding{184}
Evaluating the bandwidth and latency of common GPU {\coll}s based on NCCL Tests~\cite{nccl_tests}.
\ding{185}
Evaluating the performance of DNN training.

\noindent {$\bullet$ \textbf{Comparing Targets.}} 
We compare the bandwidth and latency of {\coll}s from {\occl} with those from NCCL~\cite{nccl}.
%
To evaluate training performance, we conducted three sets of comparative experiments:
\ding{182} In data-parallel scenarios, we compare ResNet50~\cite{he2016resnet} training throughput (\#samples consumed per second) with {\occl} versus that with NCCL orchestrated by different CPU-based methods. The comparing targets include Horovod v0.28.1, KungFu v0.2.5, and static sorting from OneFlow v0.8.1.
\ding{183} To demonstrate {\occl}'s applicability and performance under various distributed training methods, we compare the training throughput of Vision Transformer (ViT)~\cite{2021ICLR_vit} in OneFlow v0.8.1 at different scales and with different distributed training techniques.
\ding{184}
To showcase {\occl}'s performance in more popular and recent scenarios, we compare the GPT-2~\cite{radford2019gpt2, CodeParrot} training performance with {\occl} against that with manually orchestrated NCCL in PyTorch v2.2.1 \& Megatron-LM 23.06.


\subsection{{\occl}'s Deadlock-preventing Capability}
\label{sec:eval_deadlock}

We develop testing programs to demonstrate the deadlock-preventing capability of {\occl}.
The testing programs directly invoke GPU collectives and GPU synchronization operations according to the basic GPU deadlock situations at bottom library level discussed in Sec.~\ref{sec:back_basic}.

\noindent $\bullet$
In the first program, eight GPUs, each using a unique random launch order, invoke the same set of eight all-reduces with buffer sizes from 256B to 1MB.
%
Results on the 3090-server show that, without a dedicated stickiness adjustment policy, all GPUs successfully execute the eight {\occl}-based all-reduces for 200 iterations.
Approximately 18,000 preemptions occur for each block on average.

\noindent $\bullet$
In the second program, we insert \textsl{\textsf{cudaDeviceSynchronize()}} calls as GPU synchronization between all-reduces invoked in different orders on eight GPUs.
%
Results on the 3090-server show that over 200 iterations, the {\dk} on each GPU voluntarily quits for 360 times on average, ensuring successful execution of the {\occl}-based all-reduces.

NCCL's deadlock ratio is 100\% in the testing programs.

The testing programs are representative of GPU {\coll} deadlock situations that can happen during distributed DNN training, because different characteristics at application level, e.g., model type, training scale, and parallelism-type, do not introduce other basic GPU collective deadlock situations at bottom library level.

\subsection{Workload-independent Overheads}
\label{sec:eval_overheads}
%
%
Workload-independent overheads, which do not increase linearly with workload (buffer size), are divided into \textbf{memory overheads} and \textbf{time overheads}.
%
%
\ding{182}
Workload-independent memory overheads include shared memory for each block's task queue and active context slot, and global memory for the {\coll} context buffer and other related data structures.
\ding{183} Workload-independent time overheads in {\occl} include the time required for executing multiple {\coll}s within the daemon kernel, including loading and saving context.
%



{\occl} requires 13KB of shared memory and 4MB of global memory per block to maintain the block-dedicated task queue and {\coll} context buffer for 1,000 {\coll}s.
%
%
Another 11KB of global memory is needed to keep the completion counters of {\coll}s and other information accessible to all blocks.

\begin{figure}[tb]
    \centering
    \subfigure[Time Composition for a {\Coll}'s Execution in the {\DK}]{
        \begin{minipage}[t]{0.44\textwidth}
            \centering
            \includegraphics[width=\textwidth]{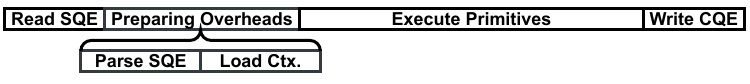}
            \label{fig:time_split_bar}
        \end{minipage}    
    }
    \\
    \subfigure[Workload-independent Time Overheads]{
        \begin{minipage}[t]{0.42\textwidth}
            \centering
            \includegraphics[width=\textwidth]{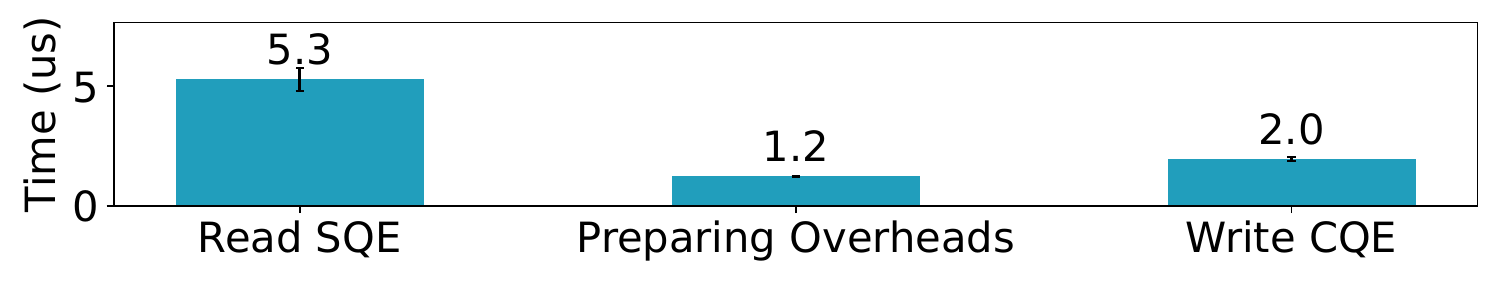}
            \label{fig:time_split_data}
        \end{minipage}    
    }
    \\
    \subfigure[Time Taken to Write CQE to Different Versions of CQ]{
        \begin{minipage}[t]{0.42\textwidth}
            \centering
            \includegraphics[width=\textwidth]{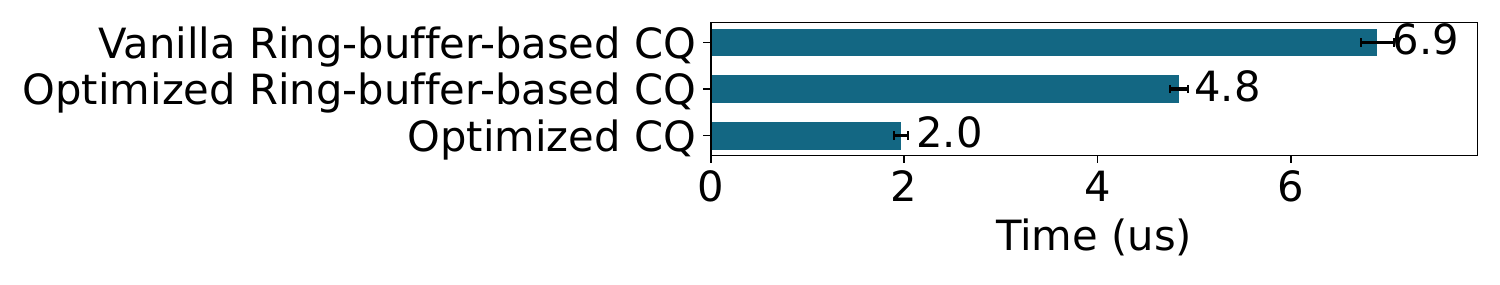}
            \label{fig:time_put_cqe}
        \end{minipage}    
    }
    \caption{Workload-independent Time Overheads Analysis.}
    \label{fig:time_split}
\end{figure}

\begin{figure*}[tb]
    \subfigure[3080ti-server, 8 GPUs, Broadcast]{
        \begin{minipage}[t]{0.32\textwidth}
            \centering
            \includegraphics[width=\textwidth]{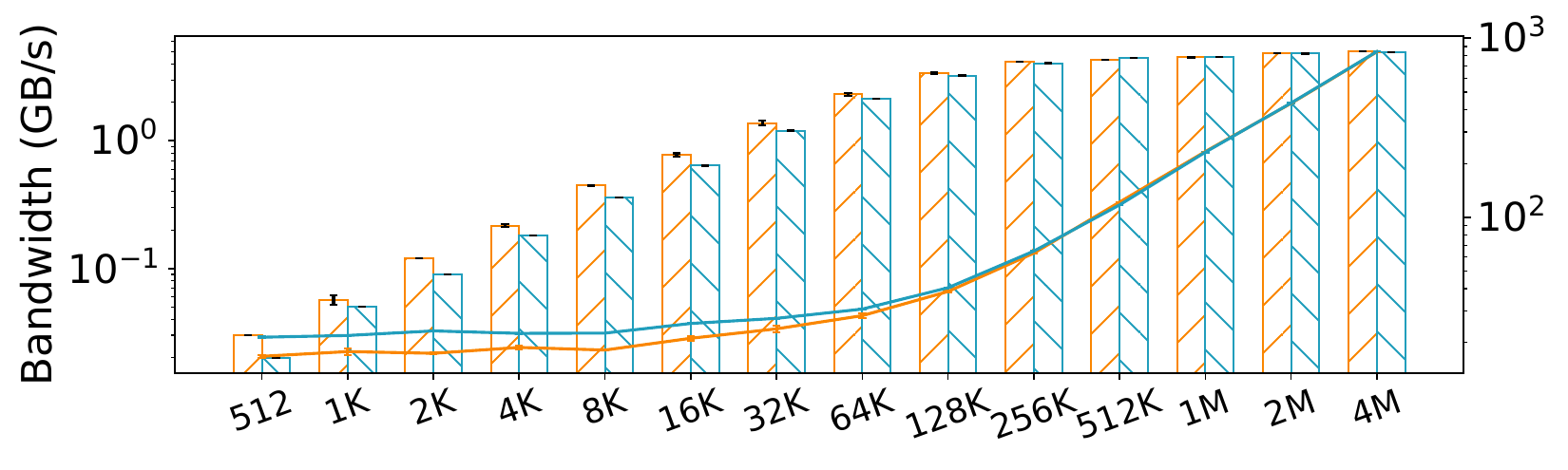}
            \label{fig:nccl_28_8card_B_main}
        \end{minipage}    
    }
    \subfigure[3090-server, 8 GPUs, All-reduce]{
        \begin{minipage}[t]{0.32\textwidth}
            \centering
            \includegraphics[width=\textwidth]{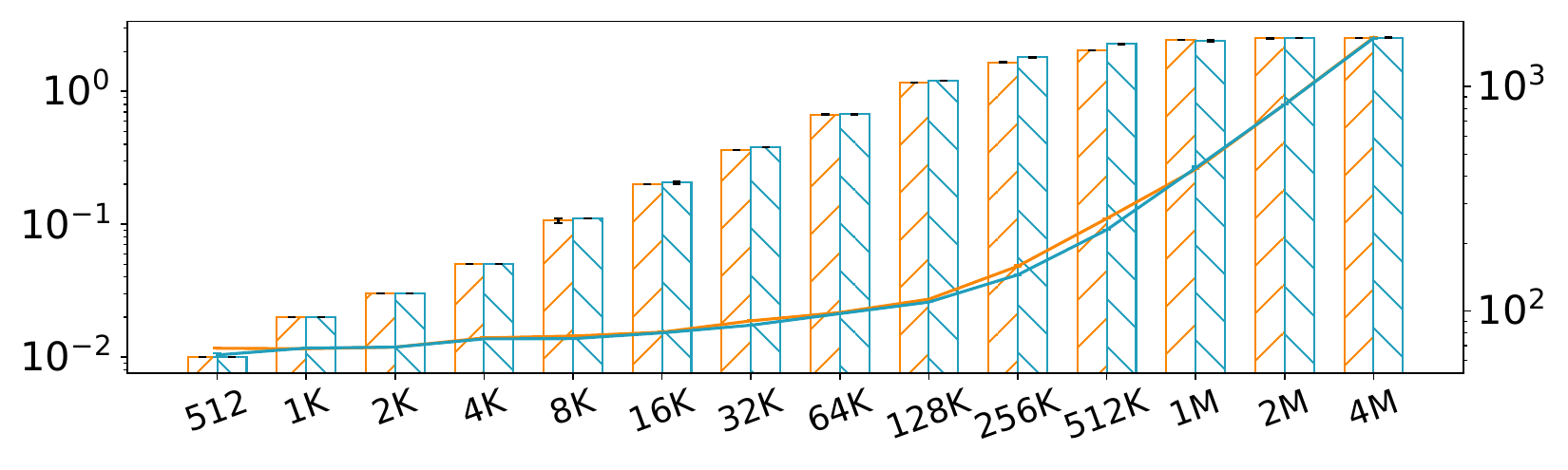}
            \label{fig:nccl_27_8card_AR_main}
        \end{minipage}    
    }
    \subfigure[2$\times$3080ti-\&2$\times$3090-server, 32 GPUs, All-reduce]{
        \begin{minipage}[t]{0.32\textwidth}
            \centering
            \includegraphics[width=\textwidth]{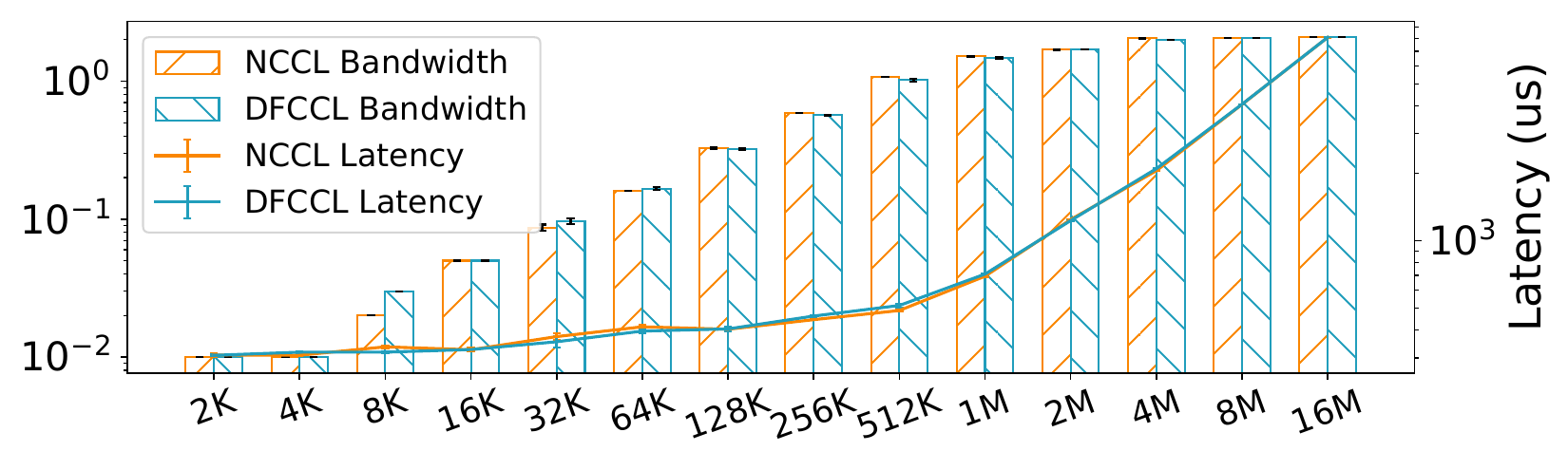}
            \label{fig:MPI_4hosts_AR_main}
        \end{minipage}    
    }
    \caption{Algorithm Bandwidth and End-to-end Latency of {\Coll}s with Different Buffer Sizes.}
    \label{fig:nccl_main}
\end{figure*}

\begin{figure}[tb]
    \subfigure[All-gather with \textbf{4KB} Buffer]{
        \begin{minipage}[t]{0.21\textwidth}
            \centering
            \includegraphics[width=\textwidth]{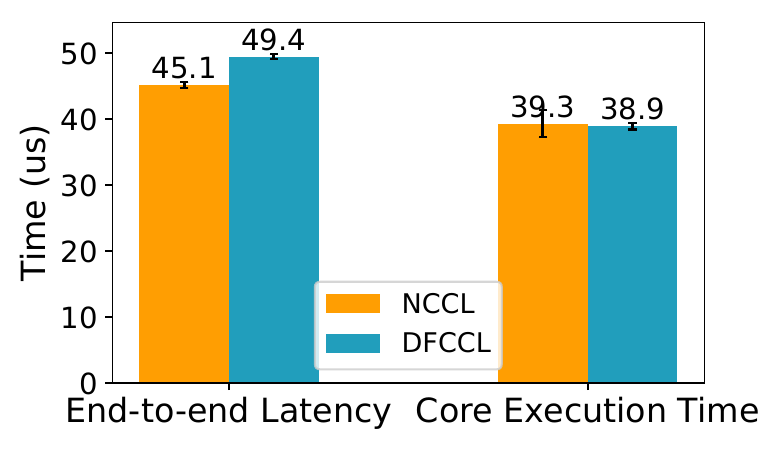}
            \label{fig:nccl_extra_AG_4K}
        \end{minipage}    
    }
    \subfigure[All-gather with \textbf{4MB} Buffer]{
        \begin{minipage}[t]{0.21\textwidth}
            \centering
            \includegraphics[width=\textwidth]{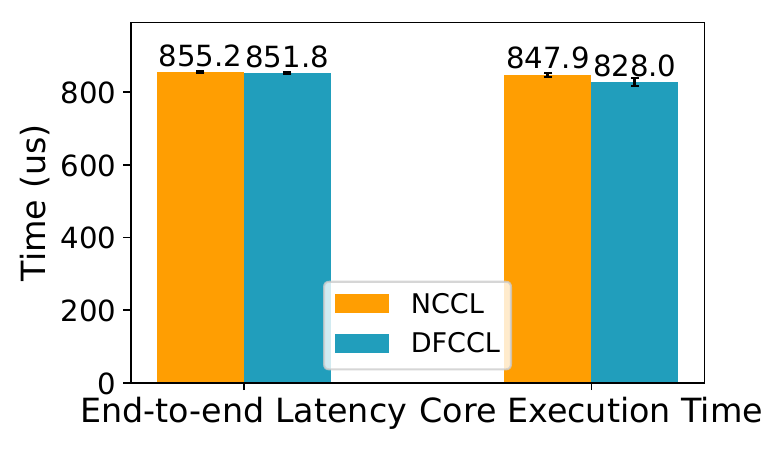}
            \label{fig:nccl_extra_RS_4M}
        \end{minipage}    
    }
    \caption{\textbf{Case Study}: Comparing End-to-end Latency and Core Execution Time of NCCL and {\occl} with Small and Large Buffer Sizes on eight GPUs of the 3090-server.}
    \label{fig:nccl_extra}
\end{figure}

Fig.~\ref{fig:time_split_bar} illustrates the timeline of executing a {\coll} in the {\dk} without preemptions.
%
%
Fig.~\ref{fig:time_split_data} illustrates the SQE read time, the preparing overheads, and the CQE write time when executing all-reduce on the 3090-server's eight GPUs.
Fig.~\ref{fig:time_put_cqe} shows the optimized CQ, detailed in Sec.~\ref{sec:implement}, reduces CQE write time from about 6.9~$\mu$s to about 2.0~$\mu$s.
The context loading takes approximately 0.45~$\mu$s, while saving context requires about 0.05~$\mu$s.

\subsection{Bandwidth and Latency of {\Coll}s}
\label{sec:eval_nccl}

We rewrite the NCCL Tests~\cite{nccl_tests} based on commit 8274cb4 to evaluate the bandwidth and latency of {\occl}-based {\coll}s, and compare {\occl} with NCCL 2.12.12.
%
Fig.~\ref{fig:nccl_main} presents selected results.
%
These results are obtained by averaging three experiments, with five iterations each.
%
%
{\occl} achieves comparable algorithm bandwidth and latency to NCCL across various scales and buffer sizes.

To further analyze {\occl}'s performance gains and overheads, we measure and compare the end-to-end latency and core execution time of {\coll}s in {\occl} and NCCL with different buffer sizes. The results are shown in Fig.~\ref{fig:nccl_extra}.
%

%
%
For {\occl}, the core execution time includes ``preparing overheads'' and ``execute primitives'' time, as shown in Fig.~\ref{fig:time_split_bar}.
For NCCL, it is the execution time of the dedicated kernel for a {\coll}.

Fig.~\ref{fig:nccl_extra_AG_4K} indicates that with a small buffer, the end-to-end latency of the {\coll} from {\occl} is about 4 $\mu$s higher than that from NCCL, yet {\occl}'s core execution time is shorter.
Meanwhile, when dealing with a larger buffer, as illustrated in Fig.~\ref{fig:nccl_extra_RS_4M}, {\occl}'s end-to-end latency is about 3~$\mu$s lower than NCCL's, and the core execution time of {\occl} is roughly 20 $\mu$s shorter than that of NCCL.

The reduced core execution time of {\coll}s in {\occl} results from the fusion of {\coll}s within the {\dk}, both spatially and temporally.
%
%
In {\occl}, the {\dk} fuses concurrently invoked {\coll}s and multiple iterations of a single {\coll}.

When a {\coll} requires fewer primitives,
its core execution time is shorter.
This makes I/O latency a more significant contributor to the {\coll}'s end-to-end latency, and renders the core execution time savings in {\occl} insufficient to offset its I/O latency, e.g., in Fig.~\ref{fig:nccl_extra_AG_4K} and when the buffer size is less than 64KB in Fig.~\ref{fig:nccl_28_8card_B_main}.
We will prioritize optimizing {\occl}'s I/O handling scheme in future work.



Nonetheless, as the workload increases, the reduction in core execution time achieved by {\occl} gradually compensates for its I/O latency, ultimately leading to a decrease in the overall end-to-end latency as shown in Fig.~\ref{fig:nccl_main}.
%
%



\subsection{DNN Training Performance}
\label{sec:eval_train}

Results in this section show that
{\occl}, using a \textit{unified} on-GPU adaptive scheduling scheme, achieves training performance on par with or exceeding NCCL, which requires \textit{different} CPU orchestration methods tailored to specific scenarios.
%
{\occl}'s efficiency also relates to its ability to execute deadlock-free collectives concurrently and asynchronously without extra CPU coordination, and low overheads for deadlock identification and context switch.
%
In the experiments, we observe different GPUs launching {\coll}s in various orders when using {\occl}, and \textit{no deadlocks} arise.
%
The loss convergence rate with {\occl} remains similar to NCCL.




\subsubsection{Data Parallel ResNet50 Training}

Experiments are conducted using eight GPUs on each of the 3080ti-server and 3090-server.
Per-GPU batch sizes are set to 48 and 96 for the 3080ti- and 3090-server, respectively. We report the average training throughput over 200 iterations.

\begin{figure}[t]
    \subfigure[3080ti-server]{
        \begin{minipage}[t]{0.21\textwidth}
            \centering
            \includegraphics[width=\textwidth]{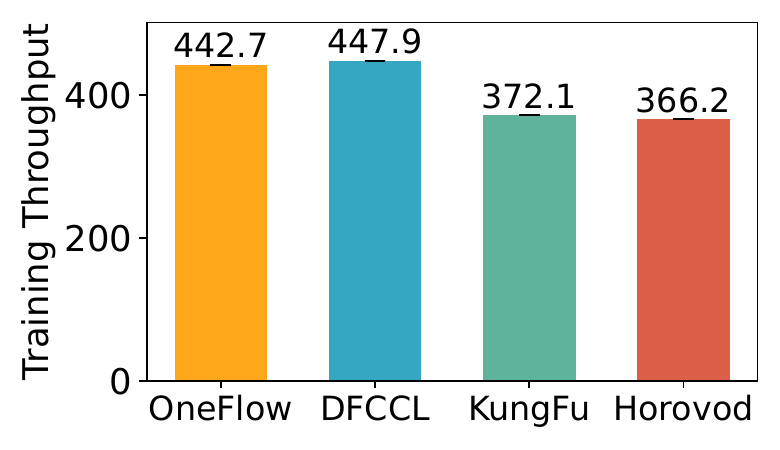}
            \label{fig:resnet_tp_3080ti_main}
        \end{minipage}    
    }
    \subfigure[3090-server]{
        \begin{minipage}[t]{0.21\textwidth}
            \centering
            \includegraphics[width=\textwidth]{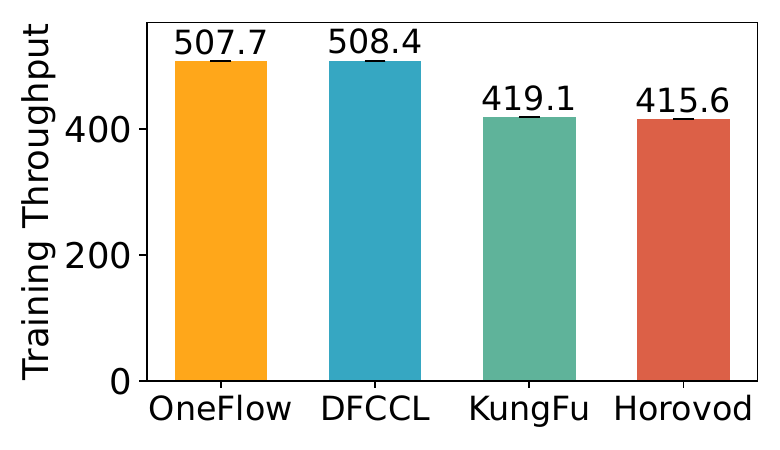}
            \label{fig:resnet_tp_3090_main}
        \end{minipage}    
    }
    \caption{Average Throughput of Training ResNet50 with Data Parallelism for 200 Iterations.}
    \label{fig:resnet_main}
\end{figure}

Fig.~\ref{fig:resnet_main} illustrates that,
the training throughput achieved with {\occl} is comparable to that attained with statically sorted NCCL in OneFlow, with improvements up to 1.2\%.
%
{\occl} outperforms KungFu and Horovod by 20.4\%-22.3\%.

\begin{figure}[t]
    \subfigure[\# Context Switches on GPU 0]{
        \begin{minipage}[t]{0.22\textwidth}
            \centering
            \includegraphics[width=\textwidth]{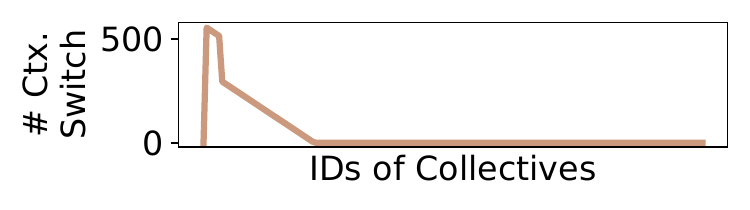}
            \label{fig:resnet_ctx_switch_GPU-0_main}
        \end{minipage}    
    }
    \subfigure[Task Queue Length on GPU 0]{
        \begin{minipage}[t]{0.22\textwidth}
            \centering
            \includegraphics[width=\textwidth]{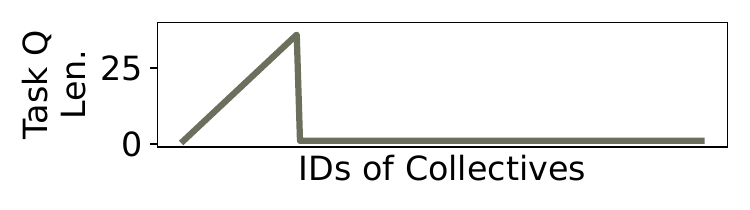}
            \label{fig:resnet_taskq_len_GPU-0_main}
        \end{minipage}    
    }
    \\
    \subfigure[\# Context Switches on GPU 2]{
        \begin{minipage}[t]{0.22\textwidth}
            \centering
            \includegraphics[width=\textwidth]{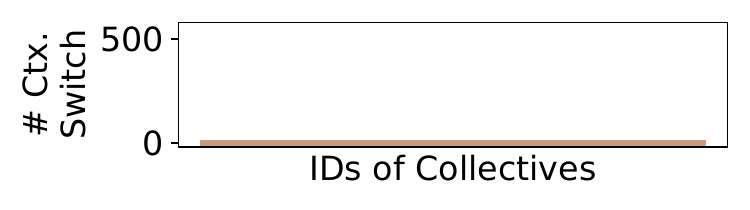}
            \label{fig:resnet_ctx_switch_GPU-2_main}
        \end{minipage}    
    }
    \subfigure[Task Queue Length on GPU 2]{
        \begin{minipage}[t]{0.22\textwidth}
            \centering
            \includegraphics[width=\textwidth]{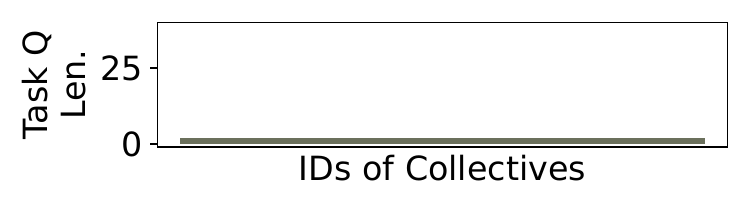}
            \label{fig:resnet_taskq_len_GPU-2_main}
        \end{minipage}    
    }
    
    \caption{Statistical Data for Assessing the Impact of the Adaptive Scheduling on Performance.}
    \label{fig:resnet_extra_main}
\end{figure}

\textbf{Case Study: Assessing the Impact of the Adaptive Scheduling on Performance.}
%
Fig.~\ref{fig:resnet_extra_main} displays two sets of statistical data from a ResNet50 training iteration with four GPUs on the 3090-server: the number of context switches and task queue lengths on different GPUs.
The X-axis of the figures in Fig.~\ref{fig:resnet_extra_main} represents IDs of the {\coll}s used in ResNet50 training.
%
These IDs are displayed in the invocation order on each GPU.
Fig.~\ref{fig:resnet_ctx_switch_GPU-0_main} and Fig.~\ref{fig:resnet_ctx_switch_GPU-2_main} display the number of context switches of each {\coll}, i.e., the number of times the {\coll} is preempted before its completion in the iteration.
%
%
Meanwhile, Fig.~\ref{fig:resnet_taskq_len_GPU-0_main} and Fig.~\ref{fig:resnet_taskq_len_GPU-2_main} illustrate the task queue length after the {\dk} reads the SQE of the corresponding {\coll}.
%
The context switch and task queue length on GPU 1 \& 3 is similar to that on GPU 0.

The spikes in Fig.~\ref{fig:resnet_extra_main} are due to a naive spin threshold adjustment policy, where each {\coll}'s spin threshold is fixed at 10,000 and does not adaptively change.
%
During the backward pass of DP, {\coll}s are invoked in bursts.
%
The scenario in Fig.~\ref{fig:resnet_extra_main} occurs when GPU 2 slightly delays issuing {\coll}s, while the other three GPUs aggressively fetch new SQEs because all fetched {\coll}s fail to progress.
As a result, {\coll}s accumulate in the task queues of the three GPUs, and {\coll}s fetched earlier are preempted multiple times.
Such a situation causes the training throughput to drop from over 500 to less than 100.


By applying the adaptive spin threshold adjusting policy, which allocates a larger initial spin threshold (100,000) via profiling for the {\coll} at the front of the task queue, and utilizes a twentyfold-greater spin threshold after primitive success, {\occl} eliminates the spike in Fig.~\ref{fig:resnet_extra_main}, and maintains high training throughput by achieving gang-scheduling.


\subsubsection{Distributed ViT Training }

\begin{figure}[tb]
    \subfigure[Data Parallelism on 8 GPUs]{
        \begin{minipage}[t]{0.225\textwidth}
            \centering
            \includegraphics[width=\textwidth]{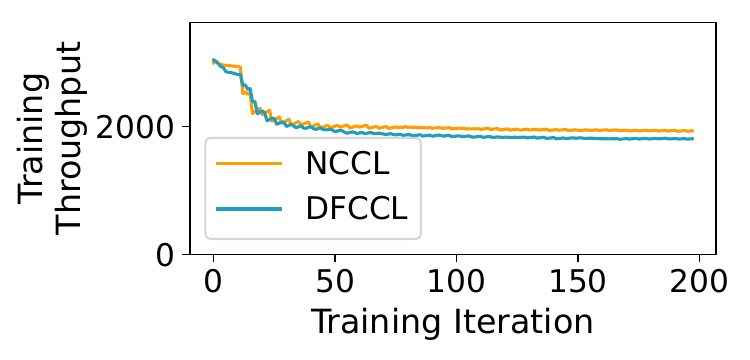}
            \label{fig:vit_dp_8card}
        \end{minipage}    
    }
    \subfigure[Tensor Parallelism on 8 GPUs]{
        \begin{minipage}[t]{0.225\textwidth}
            \centering
            \includegraphics[width=\textwidth]{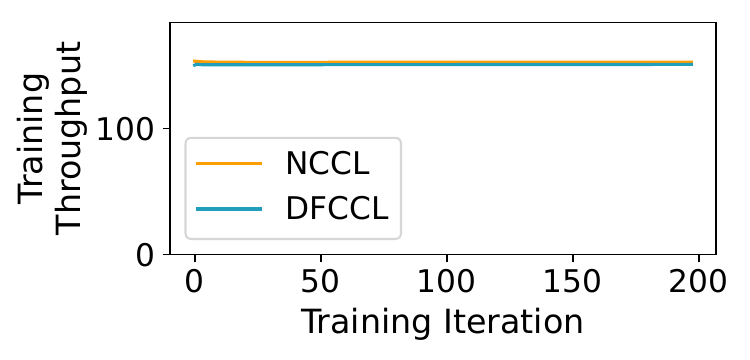}
            \label{fig:vit_tp_8card}
        \end{minipage}    
    }
    \\
    \subfigure[3D Hybrid on 16 GPUs (Base)]{
        \begin{minipage}[t]{0.225\textwidth}
            \centering
            \includegraphics[width=\textwidth]{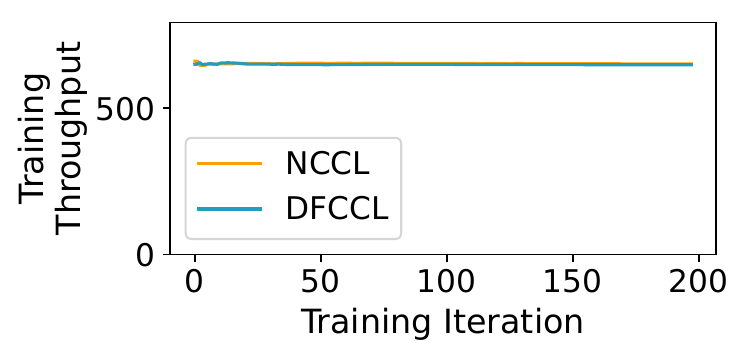}
            \label{fig:vit_16card_base}
        \end{minipage}    
    }
    \subfigure[3D Hybrid on 16 GPUs (Large)]{
        \begin{minipage}[t]{0.225\textwidth}
            \centering
            \includegraphics[width=\textwidth]{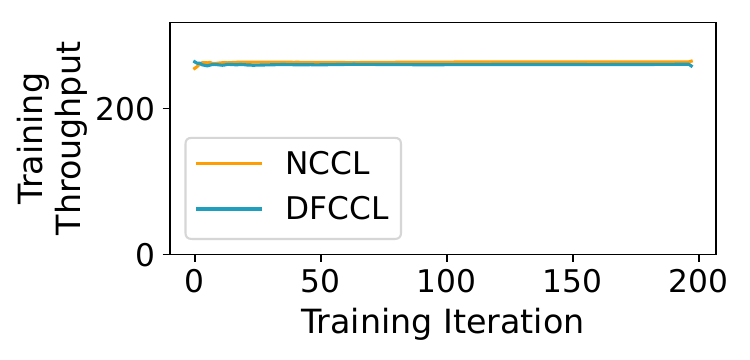}
            \label{fig:vit_16card_large}
        \end{minipage}    
    }
    \caption{Throughput for Training ViT Using OneFlow on 3090-Servers (Higher is Better).}
    \label{fig:vit}
\end{figure}
Fig.~\ref{fig:vit} shows the ViT model's training throughput under various distributed training techniques. Each iteration reports the average throughput from the start to the current iteration~\cite{wu2019detectron2}.
%
Fig.\ref{fig:vit_dp_8card}-\ref{fig:vit_16card_base} employ the base ViT configuration, while Fig.\ref{fig:vit_16card_large} utilizes the large ViT configuration.
%
The microbatch size is 128.
%
Experiments use eight GPUs on a single 3090-server (Fig.~\ref{fig:vit_dp_8card}-\ref{fig:vit_tp_8card}), and 16 GPUs across two 3090-servers (Fig.~\ref{fig:vit_16card_base}-\ref{fig:vit_16card_large}).

Results in Fig.~\ref{fig:vit} show that {\occl} efficiently provides the required {\coll}s for various parallel DNN training methods.
%
Compared to NCCL statically sorted or manually orchestrated by OneFlow, {\occl} delivers comparable performance.
%
In Fig.~\ref{fig:vit_dp_8card}, {\occl} exceeds NCCL by up to 8.6\% and maintains a gap within 7.4\%.
%
In other experiments, the difference between {\occl} and NCCL falls within ±3\%.

\begin{figure}[tb]
    \subfigure[3D Hybrid on 8 GPUs]{
        \begin{minipage}[t]{0.225\textwidth}
            \centering
            \includegraphics[width=\textwidth]{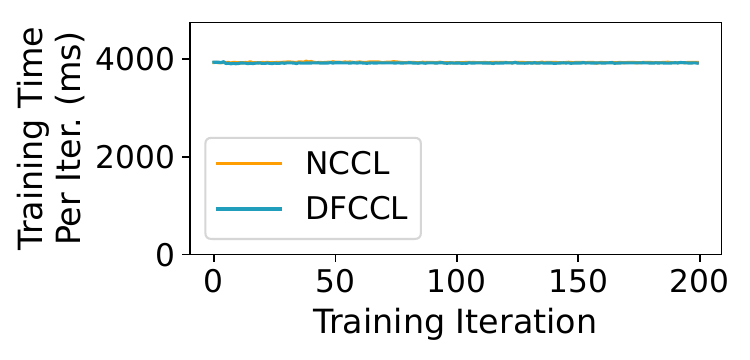}
            \label{fig:megatron_222}
        \end{minipage}    
    }
    \subfigure[3D Hybrid on 16 GPUs]{
        \begin{minipage}[t]{0.225\textwidth}
            \centering
            \includegraphics[width=\textwidth]{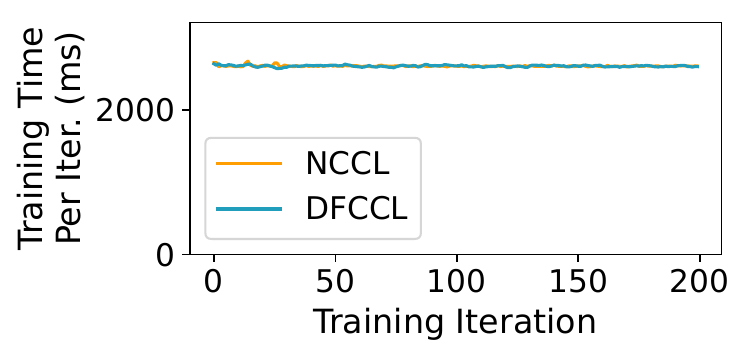}
            \label{fig:megatron_242}
        \end{minipage}    
    }
    \caption{Time Per Iteration for Training GPT-2 Using PyTorch \& Megatron-LM on 3090-Servers (Lower is Better).}
    \label{fig:megatron}
\end{figure}

\subsubsection{Distributed GPT-2 Training}

Fig.~\ref{fig:megatron} shows the per-iteration training time of GPT-2~\cite{radford2019gpt2, CodeParrot} under 3D-hybrid parallelism with Megatron-LM \& PyTorch.
%
%
The microbatch size is 18.
Results demonstrate that {\occl} provides comparable training performance to NCCL that is manually orchestrated by Megatron-LM \& PyTorch, exhibiting performance differences within ±4\%.
Furthermore, {\occl} achieves training stability comparable to NCCL. On a single 3090-server, {\occl} exhibits a 1.4\% coefficient of variation in per-iteration training time over 200 iterations, compared to NCCL's 1.5\%. When scaling to 16 GPUs across two 3090-servers, the coefficient of variation for per-iteration training time remains comparable, with 4.3\% for {\occl} and 3.9\% for NCCL.

%% file: sections/related.tex
\noindent \textbf{\textit{{\Coll} Scheduling.}}
Some {\coll} scheduling work aims to overlap communication and computation.
Poseidon~\cite{zhang2017poseidon}, TicTac~\cite{hashemi2019tictac}, P3~\cite{jayarajan2019priority_p3} are proposed in the context of parameter-server-based~\cite{limu2014ps_osdi} data parallelism, and the core insights are applied to all-reduce-based data parallelism. 
Poseidon~\cite{zhang2017poseidon} proposes wait-free backpropagation, synchronizing gradients after each layer rather than a whole iteration.
TicTac~\cite{hashemi2019tictac} adjusts gradient synchronization order so that the next iteration can launch earlier.
P3~\cite{jayarajan2019priority_p3} (priority-based parameter propagation) slices parameters into finer granularity, allowing synchronization of higher-priority parameters to preempt that of lower priority.
Bytescheduler~\cite{peng2019generic_bytescheduler} applies a similar idea to both parameter server and all-reduce architectures, and introduces an approach to auto-tuning slice size.
PACE~\cite{bao2020preemptive_pace} preemptively schedules segmented all-reduce kernels based on the directed acyclic graph (DAG) of DNN.
PyTorch Distributed~\cite{2020vldb_pyrotch_ddp} initiates all-reduce earlier than the end of the local backward pass to overlap computation with communication.
CoCoNet~\cite{jangda2022breaking_coconet} fuses split all-reduce with the computation consuming its results and overlaps the fused kernels.
{\occl} can support the above scheduling policies.


\noindent \textbf{\textit{{\Coll} Optimization.}}
Blueconnect~\cite{cho2019blueconnect} decomposes an all-reduce to many parallelizable reduce-scatters and all-gathers to adapt to network topology.
BLink~\cite{wang2020blink} proposes a heuristic spanning tree packing algorithm to optimize collective primitives.
PLink~\cite{luo2020plink} exploits the hierarchical network topology to construct a logical topology for {\coll}s.
SCCL~\cite{cai2021synthesizing_sccl} solves an integer programming encoding to achieve the Pareto-frontier of latency- and bandwidth-optimal algorithms in a single machine.
TACCL~\cite{shah2021taccl} synthesizes collective algorithms for multi-node topologies utilizing bandwidth and latency probes.
%
MSCCLang~\cite{cowan2023msccl} provides a domain-specific language describing the chunk-oriented dataflow of {\coll}s.
%
Themis~\cite{isca22_themis} dynamically assigns unique pipeline schedules to data chunks in {\coll}s to maximize utilization of all network dimensions.
%
Existing {\coll} optimization techniques utilize underlying physical topology to adjust data segmentation, chunk routing, aggregation hierarchy, etc., complementing {\occl} orthogonally.

\noindent \textbf{\textit{GPU Preemption.}}
Prior work in literature aims to support preemption for GPU to lower the end-to-end latency of high-priority tasks.
Hardware solutions~\cite{REEF70_naive, 2016SC_REEF44, Chimera_REEF56} enhance the hardware to support preemption, managing context such as registers, shared memory, barrier states, etc., during runtime.
Software methods~\cite{EffiSha_REEF12, FLEP_REEF77, GPES_REEF90, reef_39_transaction_rollback, reef_39_idempotence, Han2022REEF} can be applied directly on commodity GPUs.
Wait-based preemption approaches~\cite{EffiSha_REEF12, FLEP_REEF77, GPES_REEF90} modify user kernels to insert scheduling points so that user kernels quit more frequently and expose more scheduling chances.
Lee et al.~\cite{reef_39_transaction_rollback, reef_39_idempotence} and REEF~\cite{Han2022REEF} kill the preempted kernel directly to decrease scheduling delay.
Preempted idempotent kernels can be aborted and then restarted without affecting correctness~\cite{reef_39_idempotence, Han2022REEF}, while preempted non-idempotent kernels are rolled back and then relaunched~\cite{reef_39_transaction_rollback, reef_39_idempotence}.
{\occl} supports preemption without hardware modification.
Wait-based preemption still cannot prevent deadlocks. 
{\Coll}s are non-idempotent, and rolling back {\coll}s introduces considerable overheads and complicated synchronization issues among GPUs.
%
%
Existing GPU preemptive scheduling methods are limited to single-GPU kernels. {\occl} achieves decentralized dynamic preemption and adaptive scheduling for collectives across multiple GPUs.



%% file: sections/conclusion.tex
%
GPU {\coll} deadlocks pose threats to distributed deep learning.
We present {\occl}, a novel GPU {\coll} communication library that provides a comprehensive approach to prevent GPU {\coll} deadlocks by supporting preemption, and maintaining high performance via adaptive scheduling.
%
%
Experimental results show that {\occl} effectively prevents GPU {\coll} deadlocks and achieves performance comparable to or superior to NCCL. 
The code of {\occl} and the simulator is publicly available at \url{https://github.com/Oneflow-Inc/dfccl}.